\documentclass[12pt]{article}
\usepackage{tocloft}
\usepackage{subfiles} 
\usepackage[titletoc,toc,title]{appendix}
\usepackage[protrusion=true,expansion=true]{microtype}
\usepackage{authblk}
\usepackage[utf8]{inputenc} % allow utf-8 input
\usepackage[T1]{fontenc}    % use 8-bit T1 fonts

\usepackage{parskip,changepage,enumerate}
\usepackage[explicit]{titlesec}
\usepackage{setspace}
\usepackage[margin=0.85in]{geometry}

\usepackage{amsmath,amsthm,amsfonts,amssymb,mathrsfs,dsfont,mathtools,accents}
\usepackage{bbm}
\usepackage{etoolbox}
\usepackage{color}
\usepackage[dvipsnames]{xcolor}
\usepackage{graphicx,subfig,adjustbox} 
\usepackage{threeparttable,caption,booktabs} 
\usepackage{array,multirow,diagbox,float}
\usepackage[section]{placeins}
\usepackage[labelfont=bf]{caption}
\usepackage[font = onehalfspacing]{caption}
\usepackage{tikz}
\usetikzlibrary{positioning, fit, backgrounds, calc, patterns, arrows.meta, hobby}
\usepackage[hyperfootnotes=true]{hyperref}
\usepackage[capitalise, nameinlink]{cleveref}
\usepackage{url}            % simple URL typesetting
\usepackage{soul}

\usepackage{comment}
\usepackage{algorithm}
\usepackage{algpseudocode}
\usepackage{booktabs}
\usepackage{multirow}
\usepackage{csvsimple}
\usepackage[round,semicolon]{natbib}
\usepackage{doi}

\DeclareMathOperator{\logonep}{log1p}

\newcommand{\Ind}{\mathds{1}}
\newcommand{\CVaR}{\mathrm{CVaR}}

\setstcolor{red} 

\numberwithin{equation}{section}

\hypersetup{colorlinks = true, citecolor = MidnightBlue, linkcolor = MidnightBlue, urlcolor = MidnightBlue}

\titleformat{\section}{\normalfont\large\bfseries}{\thesection}{1em}{#1}
\titleformat{\subsection}{\normalfont\normalsize\bfseries}{\thesubsection}{1em}{#1}
\titleformat{\subsubsection}{\normalfont\normalsize\itshape}{\thesubsubsection}{1em}{#1}
\titlespacing\section{0pt}{12pt plus 4pt minus 2pt}{6pt plus 2pt minus 2pt}
\titlespacing\subsection{0pt}{12pt plus 4pt minus 2pt}{3pt plus 2pt minus 3pt}
\titlespacing\subsubsection{0pt}{12pt plus 4pt minus 2pt}{0pt plus 2pt minus 3pt}

\crefname{appendix}{Appendix}{Appendices}
\Crefname{appendix}{Appendix}{Appendices}

\newtheorem{corollary}{Corollary}[section]
\newtheorem{definition}{Definition}[section]

\def\boxit#1{\vbox{\hrule\hbox{\vrule\kern6pt
          \vbox{\kern6pt#1\kern6pt}\kern6pt\vrule}\hrule}}

\definecolor{myorange}{rgb}{1,0.5,0}
\definecolor{MyDarkBlue}{rgb}{0,0.08,0.45}

\def\fredcomment#1{\vskip 2mm\boxit{\vskip 2mm{\color{MyDarkBlue}\bf#1}
{\color{MyDarkBlue}\bf -- Fred\vskip 2mm}}\vskip 2mm}
\def\Shuyicomment #1{\vskip 2mm\boxit{\vskip 2mm{\color{orange}\bf#1}
{\color{orange}\bf -- Shuyi\vskip 2mm}}\vskip 2mm}
\begin{document}

\title{\Large \bfseries 
Insights on Time-consistent Deep Hedging under Elicitable Dynamic Risk Measures
}

\author[a, b]{Shuyi Zhang}

\author[c]{Fr\'ed\'eric Godin}

\affil[a]{{\small
River Hill High School, Intern/Mentor Research Program, Clarksville, Maryland
}}

\affil[b]{{\small
\textit{Present address:} University of Maryland, Department of Computer Science, College Park, Maryland\\
\texttt{szhang48@terpmail.umd.edu}
}}

\affil[c]{{\small
Concordia University, Department of Mathematics and Statistics, Montréal, Canada\\
\texttt{frederic.godin@concordia.ca}
}}

\vspace{-10pt}
\date{\today}

%%%%%%%%%%%% ABSTRACT PAGE %%%%%%%%%%%%%%%%%%

\maketitle \thispagestyle{empty} %\vfill \pagebreak \vspace*{\fill}

\vspace{-15pt}

\begin{abstract}
\vspace{-5pt}
We study deep hedging where the objective is expressed in terms of a dynamic risk measure, under which sequential decisions are time-consistent. We tackle the problem of basket option hedging. We show that the approach is feasible and can be used conveniently in the presence of more complex state spaces with higher dimension than what was previously test in the literature. We rely on the conditional elicitability of spectral risk measures to represent the optimization objective. We provide insights on how the choice of scoring function impacts the training of the hedging agent. We also propose an adapted reinforcement learning (RL) algorithm to solve the problem, which contains additions that are crucial to a successful training of the deep RL agent, and without which the agent fails to converge to proper solutions. The time-consistent hedging strategies are also benchmarked against deep hedging approaches relying on static risk measures leading to precommitment.

\bigskip 

\noindent \textbf{JEL classification:} C45, C61, G13, G32

\vspace{0.08in}

\noindent \textbf{Keywords:} Deep Reinforcement Learning, Actor-critic, Stochastic Control, Dynamic Risk Measures, Deep Hedging, Elicitability, Scoring Functions, Time-consistency, Basket Options
\end{abstract}

\medskip

\thispagestyle{empty} \vfill \pagebreak

%%%%%%%%%%%%%%%%%%%%%%%%%%%%%%%%%%%%%%%%%%%%%%%%

\section{Intro}
\label{sec:DHRL}
Reinforcement learning (RL) is a popular method used to solve stochastic control optimization problems~\citep{BartoSuttonAnderson1983}. Notably, RL has surpassed human performance in Go~\citep{SilverEtAl2016}, Atari $2600$~\citep{MnihEtAl2015}, and chess~\citep{SilverEtAl2018}, largely due to its ability to develop sequential decision making strategies in uncertain environments. Recently, RL has been applied to financial problems, with a strong emphasis on risk management dependent on market and portfolio dynamics. A particular avenue of RL applied to finance is dynamic hedging, where one mitigates the risk of some financial derivative by trading the underlying asset(s) the financial derivative is based upon. \citet{Buehler2019} and \cite{Halperin2020} were the first to apply deep reinforcement learning (DRL), the fusion of RL algorithms and deep neural networks, to dynamic hedging. Following these seminal works, a growing body of literature has applied DRL to increasingly complex hedging environments, incorporating features such as market frictions, more realistic market dynamics, richer state spaces, larger portfolios, more complex derivatives, and additional hedging instruments~\citep{CaoEtAl2023, CarbonneauGodin2024EqualRisk, ChenEtAl2024, DingEtAl2025, DuEtAl2020, FrancoisEtAl2024, FrancoisEtAl2025, lutkebohmert2022robust, MarzbanDelageLi2023, MikkilaKanniainen2023, MurrayEtAl2022, NeaguEtAl2024, NeaguEtAl2025, SharmaEtAl2024, wu2023robust}. 

In these contexts, typical RL frameworks involve optimizing a global objective using a \emph{static} risk measure. This refers to the risk associated with the overall distribution of the cumulative reward, consistent with the broader risk-aware RL literature~\citep{PrashanthGhavamzadeh2016, TamarMannor2013}. However, these frameworks often result in policies that are suboptimal over time. To resolve this, a common design choice of RL frameworks is that of \emph{time-consistency}, a property that ensures that policies remain optimal across time~\citep{CoacheJaimungal2024, YuXYing2023}. RL algorithms are particularly favorable when such frameworks are used within continuous (and therefore infinite) state and action spaces settings because they are much more efficient than conventional dynamic programming (DP) techniques, as the latter requires looping over all possible outcomes to find the optimal solution. At the same time, time-consistent risk-aware RL frameworks often rely on using \emph{dynamic} risk measures, which evaluate future risk conditional on the information available at each point in time, to achieve time-consistency, but such approaches may suffer from the burden of recursively evaluating the dynamic risk measure through nested simulations. 

A recent development is the integration of elicitability into risk-aware RL frameworks, allowing dynamic risk measures to be estimated efficiently by minimizing scoring (loss) functions without the computational burden of nested simulations~\citep{CoacheJaimungalCartea2023, MarzbanDelageLi2023}. While these approaches alleviate the computational challenges, there remains opportunity to further explore how elicitable risk frameworks can be designed in deep learning settings to produce desirable optimization landscapes for risk estimation and facilitate more effective convergence toward optimal RL policies in high-dimensional settings. In this paper, we analyze different characterizations of strictly consistent scoring functions and use these insights to develop an actor-critic RL framework that leverages the conditional elicitability of dynamic spectral risk measures, employs a stochastic policy, and targets desirable optimization geometry by exploiting the translation invariance of spectral risk measures. We characterize the optimization properties of different scoring functions and empirically examine their effects on the performance of risk-aware policies derived from the proposed algorithm in a finite-horizon setting with a high-dimensional state space and granular temporal resolution. We also propose to include a shifting procedure within the reinforcement learning algorithm, where translations are applied to the value function estimate during training. Because the risk measures considered are translation invariant, the shifted value functions do not impact the optimal policies, but they allow to leverage regions of the score functions with higher convexity, which in turns helps stabilizing training. In our tests, failure to apply this shifting approach leads to unstable or divergent training results.

The risk-aware RL setting considered in this paper is a dynamic hedging problem involving the mitigation of risk for a portfolio containing a European basket option with a one-year maturity through the daily trading of its underlying assets. We consider both a time-consistent (dynamic) CVaR objective, for which the conditional elicitability framework is employed, and a precommitment (static) CVaR objective as a benchmark. The underlying assets are modeled according to the Dynamic Conditional Correlation - Generalized Autoregressive Conditional Heteroskedasticity (DCC-GARCH(1, 1)) framework. For our numerical experiments, we demonstrate that the proposed actor-critic RL framework under conditional elicitability can derive effective dynamic risk hedging policies in higher-dimensional settings, an aspect that has received limited attention in the existing literature. Our findings are as follows: (1) We find that high tail-sensitivity scoring functions such as the canonical logarithmic or fractional-power characterizations yield RL-based dynamic risk estimates that are the most accurate relative to critics trained using a nested approach benchmark when sufficient extreme-tail samples are available at high confidence levels. Conversely, scoring functions producing saturating gradients can provide enhanced performance and improved stability for extreme risk levels for which tail observations are very sparse. (2) While a precommitment static risk objective is the optimal choice when mitigating terminal hedging risk of basket options, deploying a dynamic risk objective yields a hedging policy with consistently lower hedging risk over shorter horizons, and a monotonic decrease in risk as maturity shortens. This is in contrast to results from the static risk policy, whose hedging risk increases as maturity decreases. This motivates the use of time-consistent risk objectives for hedging policies.
%\Shuyicomment{Since the experiments have now been completed, I added the concrete findings to this paragraph.}
The paper is organized as follows. First, we outline the elicitable, time-consistent dynamic risk framework and the theoretical considerations when characterizing scoring functions in Section~\ref{sec:EDR}. We then propose the main actor-critic algorithm in Section~\ref{sec:ACRL}, and the underlying basket hedging problem it aims to solve in Section~\ref{sec:ExpMeth}. Experimental results are provided and further discussion is carried out in Section~\ref{sec:Expres}.\footnote{The python code to replicate the experiment can be found at \url{https://github.com/shuyizhang01/Time-consistent_Deep-hedging}.} Finally, we conclude in Section~\ref{sec:conclusion}.

%\fredcomment{This is the list of identified contributions. (1) Show the coache approach is feasible in higher dimension, (2) compare to precommitted strategies, (3) assess the impact of choice of scoring function and of offsetting/shifting approach for enhancing optimization, (4) assess whether the approximation of dynamic risk by the neural approach is good and close to the nested heuristic.}

 %%%%%%%%%%%%%%%%%%%%%%%%%%%
\section{Conditionally Elicitable Time-consistent Dynamic Risk}
\label{sec:EDR}
Following ideas from \cite{Ruszczynski2010} and \citet{CoacheJaimungalCartea2023}, we outline the theoretical setup we consider for risk optimization under time-consistent elicitable dynamic risk measures.

\subsection{Time-consistent dynamic risk measures}

We first outline the structure of risk measures used in this paper. 

Let $\mathcal{T} := \{0, \dots, T\}$ be a sequence of time periods and
$(\Omega, \mathcal{F}, \{\mathcal{F}_t\}_{t \in \mathcal{T}}, \mathbb{P})$
a filtered probability space satisfying the usual conditions.
Define $\mathcal{X}_t$ as the set of $\mathcal{F}_t$-measurable random variables for $t \in \mathcal{T}$. We consider a sequence of one-step conditional risk measures $\{\rho_t\}^{T-1}_{t=0}$ with each mapping $\rho_{t} : \mathcal{X}_{t+1} \rightarrow \mathcal{X}_{t}$ possessing the following properties:
\begin{itemize}
    \item \textbf{Monotonicity:} For any $X, Y \in \mathcal{X}_{t+1}$ with $X \geq Y$, $\rho_t(X) \geq \rho_t(Y)$.
    \item \textbf{Translation Invariance:} For any $X\in \mathcal{X}_{t+1}$ and $m \in \mathbb{R}$, $\rho_t(X + m) = \rho_t(X) + m$.
    \item \textbf{Null at Zero Risk:} $\rho_t(0) = 0$.
    \item \textbf{Law invariance:} If $X \in \mathcal{X}_{t+1}$ and $Y \in \mathcal{X}_{t+1}$ have the same conditional distribution with respect to (w.r.t.) $\mathcal{F}_t$, then $\rho_t(X) = \rho_t(Y)$.
\end{itemize}
\cite{Ruszczynski2010} highlights that such one-step conditional risk measures can be leveraged to design so-called dynamic risk measures $\{\rho_{t, T}\}^T_{t=0}$. In our work, for a $\mathcal{F}$-adapted stochastic process $\{X_t\}^T_{t=0}$, we consider dynamic risk measures of the form
\begin{align}
    \rho_{t, T}(X_t,X_{t+1}\ldots,X_T) &\equiv X_t + \rho_{t}(X_{t+1} + \rho_{t+1}(X_{t+2} + \rho_{t+2}(X_{t+3} + \ldots +\rho_{T-1}( X_T))\dots)) \label{nestRMdef}
    \\ &= \rho_t \circ \rho_{t+1} \circ \cdots \circ \rho_{T-1} \left( \sum^T_{k=t} X_k \right). \notag
\end{align}
Risk measures defined through such nesting approach possess favorable properties, such as that of time-consistency. A dynamic risk measure $\{\rho_{t, T}\}^T_{t=0}$ is said to be time-consistent if, for any $\mathcal{F}$-adapted processes $\{X_t\}^T_{t=0}$ and $\{Y_t\}^T_{t=0}$ and any $0 \leq \tau_1 < \tau_2 \leq T$, conditions $X_{\tau_1} = Y_{\tau_1}, \ldots, X_{\tau_2-1} = Y_{\tau_2-1}$ and $\rho_{\tau_2,T}(X_{\tau_2},\ldots, X_T) \geq \rho_{\tau_2,T}(Y_{\tau_2},\ldots, Y_T)$ imply that $\rho_{\tau_1,T}(X_{\tau_1},\ldots, X_T) \geq \rho_{\tau_1,T}(Y_{\tau_1},\ldots, Y_T)$. The interpretation of such property is that relative preference between risks persists through time when cycling through elements $\rho_{t, T}$ of the dynamic risk measure.
%. Indeed, defining for any integers $t_1 < t_2 <T$ 
%\begin{equation*}
    %\rho_{t_1, t_2}(X_{t_1},\ldots, X_{t_{2}}) \equiv \rho_{t_1, T}(X_{t_1},\ldots, X_{t_{2}}, 0,%\ldots,0),
%\end{equation*}

One can then show that
\begin{equation}
\label{RecurExpress}
    \rho_{t, T}(X_{t},\ldots, X_{T}) = X_t + \rho_t(\rho_{t+1, T}(X_{t+1},\ldots,X_T))
\end{equation}
which entails that the risk-to-go can be computed through a recursion analogous to the Bellman equation. 
\subsection{Integration in Markov decision processes and neural approximations}
In RL frameworks, an agent interacts with some environment and learns a policy over time. Accordingly, we consider a \emph{Markov decision process} (MDP) with finite horizon $\mathcal{T}$ defined by the tuple $(\mathcal{S}, \mathcal{A}, \mathcal{P}, \mathcal{C})$ containing a continuous state and action space, stationary transition probabilities, and a cost space. Because we are working in a finite horizon, we assume that the current stage $t$ is an element of the state vector $s_t \in \mathcal{S}$. The symbol $\pi$ denotes a stochastic policy, which associates, for any state vector $s_t$, a probability distribution over actions $a_t \in \mathcal{A}$. Thus, for each time period $t$, state variables $s_t$ are observed, an action $a_t$ is then drawn based on $\pi$, and transition probabilities characterize the next state and policy-induced $\mathcal{F}_{t+1}$-measurable cost $c_{t+1} \in \mathcal{C}$ observed:\footnote{Here $x_{0:t}$ denotes $x_0,\ldots x_t$.}
\begin{equation*}
    \mathbb{P}(s_{t+1}=s', c_{t+1} = c\mid a_t=a, s_t=s, a_{0:t-1}, s_{0:t-1})=\mathcal{P}(s^{'}, c \mid s, a).
\end{equation*}
%$\mathbb{P}$ denotes stationary transition probabilities $\mathbb{P}(s^{'} \mid s, a)$, with one-step joint distribution over state and action $\mathbb{P}(s_{t+1}, a_t \mid s_t) = \mathbb{P}(s_{t+1} \mid s_t, a_t)\pi^{\theta}(a_t \mid s_t)$. For each time $t \in \mathcal{T}$, the policy $\pi^{\theta}$ receives $s_t$, outputs action $a_t$, transitions to the next state $s_{t+1}$ with transition probability $\mathbb{P}(s_{t+1} \mid s_t, a_t)$, and induces an $\mathcal{F}_{t+1}$-measurable bounded cost $c_t$.
In this framework, setting $c^\pi_0\equiv0$, we solve for the policy that minimizes the initial risk-to-go:
\begin{equation}
\pi^* \equiv \underset{\pi}{\arg\!\min} \, \rho_{0,T}(c^\pi_0,\ldots, c^\pi_{T} ),
\label{eq:nestedrisk}
\end{equation}
with any of the optimal policies being selected in case of non-uniqueness, and where the dependence of costs $c$ on the policy $\pi$ is made explicit.
Due to time-consistency, the policy remains optimal at subsequent stages: for all $t\in \mathcal{T}$,
\begin{equation*}
    \label{eq:remainoptim}
    \pi^* \in \underset{\pi}{\arg\!\min} \, \rho_{t,T}(c^\pi_t,\ldots, c^\pi_T ).
\end{equation*}
Furthermore, the Markovian nature of the evolution of the state vector combined with the recursive expression \eqref{RecurExpress} allows defining a value function which reflects the risk-to-go. For a policy $\pi$, define
\begin{equation}
    V^\pi_t(s_t) \equiv \rho_{t,T}( (0,c^\pi_{t+1:T}) ) = \rho_t(c^\pi_{t+1} + V^\pi_{t+1}(s_{t+1})), \quad t<T,\label{eq:value function-nested}
\end{equation}
with $V^\pi_{T}(s_{T}))=0$.
The optimal value function $V^*$ associated with an optimal policy $\pi^*$ should thus satisfy
\begin{equation}
    V^*_t(s_t) = \min_{\pi} V^\pi_t(s_t) =  \min_{a_t} \, \left[\rho_t(c_{t+1} +  V^{*}_{t+1}(s_{t+1})) \right], \quad t<T. \label{eq:optimvalue function-nested}
\end{equation}

\subsection{Conditional elicitability}
\label{sec:ERM}

For one-step conditional risk measures $\rho_t$, we use (conditional) spectral risk measures operators as they admit an elicitability structure that enables efficient characterization of time-consistent dynamic risk and yield an explicit gradient formula for the value-function as highlighted in \cite{CoacheJaimungalCartea2023}.
%\cite{CoacheJaimungalCartea2023} show that the use of spectral risk measures enables the use of joint elicitability for time-consistent dynamic risk estimation. 
%This is crucial because 
Indeed, elicitability means that risk can be characterized by minimizing some scoring (loss) function which eliminates the need for nested simulation to evaluate risk. This drastically increases the efficiency of the training process when solving the hedging optimization problem. 

A static risk measure is a mapping from some random variable to a scalar risk value. Such mapping is said to be elicitable if it can be recovered as the minimizer of an expected scoring function~\citep{lambert2008eliciting, Gneiting2011}. 
\begin{definition}[Elicitability]
A statistical functional $\mathcal{R}: \mathbb{F} \to \mathbb{R}^p$, mapping a class of probability distributions $\mathbb{F}$ to the (multivariate) reals, is \textbf{elicitable}\footnote{When $p>1$, we refer to joint elicitability.} if there exists a subset $\mathbb{A} \subseteq \mathbb{R}^p$ and scoring 
function $S: \mathbb{A} \times \mathbb{R} \to \mathbb{R}$ such that
\begin{equation}
    \mathcal{R}(F) = \underset{\mathfrak{a} \in \mathbb{A}}{\arg\min} \; \mathbb{E}_{Y \sim F}[S(\mathfrak{a}, Y)], 
    \quad \forall F \in \mathbb{F}. \label{elicimapping}
\end{equation}
The scoring function $S$ is called a consistent scoring 
function for $\mathcal{R}$ if the above holds, and is said to be strictly consistent if the minimizer is unique.
\end{definition}

\begin{comment}

Denote by $\mathcal{Y}$ the set of random variables having support on $\mathbb{Y}$, and let  a cumulative distribution function (CDF) $F \in \mathbb{F}$. Consider a scoring function $
S : \mathbb{A} \times \mathbb{Y} \to \mathbb{R},
$ that returns a scalar quantity for some $\mathfrak{a} \in \mathbb{A} \subseteq \mathbb{R} $ and $y \in \mathbb{Y}$. A mapping $T : \mathbb{Y} \to \mathbb{A}$ is elicitable if
\begin{equation}
\label{elicimapping}
T(Y)
\;=\;
\arg\min_{\mathfrak{a} \in \mathbb{A}} \int S(\mathfrak{a}, y)F(\mathrm{d}y).
\end{equation}
Additionally, if $T(Y)$ is the unique minimizer of the optimization problem \eqref{elicimapping}, the scoring function $S$ is said to be strictly $\mathbb{F}$-consistent.
\end{comment}

%scores are often designed to be strictly-consistent such that the score is uniquely minimized at the true risk values.
%\begin{definition}
%A scoring rule $S:\mathbb{A}\times\mathbb{Y}\to\mathbb{R}$ is \textbf{strictly $\mathbb{F}$-consistent} for a mapping $T$ if, for every $F\in\mathbb{F}$, $T(Y)$ is the unique solution to
%\begin{equation}
 %   \arg \min_{\mathfrak{a} \in \mathbb{A}} \int S(\mathfrak{a}, y)\, F(\mathrm{d}y).
%\end{equation}
%\end{definition}
We focus on the one-step conditional dynamic CVaR risk measure which operates on one-step costs in the RL environment. In contrast to the static CVaR risk measure, which minimizes the tail risk of the cumulative reward, dynamic CVaR controls tail risk conditionally throughout the trajectory, allowing the hedging agent to adapt its actions to the level of risk associated with each state. To construct a procedure that assesses the dynamic risk of some fixed policy, we require estimation of the one-step conditional risk at each time period and state in the environment. We obtain these estimates by exploiting the joint elicitability of CVaR and VaR, as described below. First, for any random variable $Y$, define the VaR as
\begin{equation}
\mathrm{VaR}_{\alpha}(Y) = \inf\{x \mid F_Y(x) \geq \alpha\}.
\end{equation}
%with $F_Y(\dot |\mathcal{F}_t)$ representing the time-$t$ conditional CDF of $Y$.
Then, the CVaR at the $\alpha \in (0, 1) $ confidence level averages 
the worst $(1-\alpha)$ outcomes:
\begin{equation}
\mathrm{CVaR}_{\alpha} = \frac{1}{1-\alpha}\int_{\alpha}^{1}\mathrm{VaR}_{u}du.
\end{equation}
We thus consider eliciting the vector-valued mapping
\begin{align} \label{eq:Tdef}
    \mathcal{R}(Y) = (\mathrm{VaR}_{\alpha}(Y), \mathrm{CVaR}_{\alpha}(Y)),
\end{align}
%we let .
%To connect this static elicitability framework to that of dynamic risk, we let $\mathbb{F}$ denote the class of conditional and continuous CDFs of $Y \mid \mathcal{F}_t$, for $Y \in \mathcal{Y}_{t+1}$, each having a finite first moment and support on $\mathbb{Y} \subseteq \mathbb{R}$. 
and utilize the strictly consistent scoring function defined in Corollary $4.4$ of~\citet{CoacheJaimungalCartea2023} for this purpose, since it is easy to integrate in RL environments: 
\begin{equation}
\begin{split}
S(\mathfrak{a}_1, \mathfrak{a}_2, y) = 
\left(\Ind_{y \leq \mathfrak{a}_1} - \alpha\right)\left(H(\mathfrak{a}_1) - H(y)\right)
 - G(\mathfrak{a}_2) + G(y) \\
+ G'(\mathfrak{a}_2)\left[\mathfrak{a}_2 + \frac{1}{1-\alpha}\left(\mathfrak{a}_1\left(\Ind_{y > \mathfrak{a}_1} - (1-\alpha)\right) - y\Ind_{y > \mathfrak{a}_1}\right)\right]
\end{split}
\label{eq:CVaRScore}
\end{equation}
for $\mathbb{A} = \{\mathfrak{a}=(\mathfrak{a}_1,\mathfrak{a}_2) \in \mathbb{R} \times \mathbb{R} \mid \mathfrak{a}_1 \leq \mathfrak{a}_2\}$ and where $G$ is any function such that
\begin{enumerate}
    \item $\mathbb{E}_{Y \sim F}\left[|H(Y)|\right], \mathbb{E}_{Y \sim F}\left[|G(Y)|\right] < \infty$ for all $F \in \mathbb{F}$,
    \item $G: \mathbb{R} \to \mathbb{R}$ is strictly convex,
    \item The gradient of $G$, denoted $G'$, exists and for a specific $H : \mathbb{R} \rightarrow \mathbb{R}$ 
\begin{equation}
    z \mapsto H(z) -\frac{z}{1-\alpha}G'(\mathfrak{a}_2)
\label{eq:Affinemap}
\end{equation}
is strictly increasing for any $\mathfrak{a}_2 \in \mathbb{R}$.
\end{enumerate}
%In our RL experiments, we assume that $Y$ is a random variable that takes both positive and negative values, with $Y > -C$ for some constant $C > 0$ and bounded above by a finite constant.
%\fredcomment{Are gains/losses really bounded in our experiments?}
%\Shuyicomment{They are bounded by some lower value in the sense that the score clamps them at -C, though they are techincally not bounded by a larger value; just constrained to the accumulation of risk within each group. I believe that sentence was added just to give the reader an understanding that our state space and costs are sufficiently compact so as to be able to derive convergent policies, though you can edit it if desired.}
%This means that $\mathfrak{a}_1$ and $\mathfrak{a}_2$ are restricted to a bounded domain. 
To further simplify \eqref{eq:CVaRScore}, we consider the case of a null-valued function $H$, which leads to 
\begin{equation}
\begin{split}
S(\mathfrak{a}_1, \mathfrak{a}_2, y) = 
 - G(\mathfrak{a}_2) + G(y) + G'(\mathfrak{a}_2)\left[\mathfrak{a}_2 + \frac{1}{1-\alpha}\left(\mathfrak{a}_1\left(\Ind_{y > \mathfrak{a}_1} - (1-\alpha)\right) - y\Ind_{y > \mathfrak{a}_1}\right)\right].
\end{split}
\label{eq:CVaRScore2}
\end{equation}
The use of \eqref{eq:CVaRScore2} still provides great flexibility by allowing for various possible choices of $G$; testing for the impact of $G$ on the training performance of the hedging agent is a key contribution of our work. 

We can estimate the value function for all $t \in \mathcal{T}$ and $s_t \in \mathcal{S}$ in the style of a quantile regression problem~\citep{Meinshausen2006} by approximating the conditional risk mapping through elicitability. Set $\rho_t$ as the conditional CVaR with confidence level $\alpha$ defined through
\begin{equation}
    \mathrm{CVaR}_{t,\alpha} = \frac{1}{1-\alpha}\int_{\alpha}^{1}\mathrm{VaR}_{t,u}du,
    \quad \mathrm{VaR}_{t,\alpha}(Y) = \inf\{x \mid F_Y(x |\mathcal{F}_t) \geq \alpha\}.
\end{equation}
with $F_Y( \cdot |\mathcal{F}_t)$ being the time-$t$ conditional CDF of $Y$. Since our environment is Markov, we may assume that $s_t$ sufficiently summarizes the information contained in $\mathcal{F}_t$. Therefore, using \eqref{eq:optimvalue function-nested}, \eqref{elicimapping} and \eqref{eq:Tdef},
\begin{equation*}
    V^*_t(s_t) =  \min_{a_t} Proj_2 \left\{\underset{\mathfrak{a}_1, \mathfrak{a}_2}{\arg\!\min} \, \mathbb{E} \left[S(\mathfrak{a}_1,\mathfrak{a}_2,  c_{t+1} + V^{*}_{t+1}(s_{t+1})) | s_t\right] \right\} . %\label{eq:optimvalue function-nested}
\end{equation*}
where $Proj_2(\mathfrak{a}_1,\mathfrak{a}_2) =\mathfrak{a}_2$.

%The random variable $Y \mid \mathcal{F}_t$  can be represented through some state $s \in S$ with features describing the filtration $\mathcal{F}_t$. Parameterizing the value-function $V^{\Psi}$ as a neural network with parameters $\Psi$, its estimation is achieved by minimizing the expected scores for all conditional risk estimates such that
%\begin{equation}
    %V^{\bar{\Psi}} = T(Y \mid S=s) = \arg \min \limits_{V^{\Psi}} \mathbb{E}_{Y \sim %F}\left[S(V^{\Psi}(s), Y)\right].
%    \label{eq:estimateVargmin}
%\end{equation}
%where $\bar{\Psi}$ represents the optimal parameters. 

%%%%%%%%%%%%%%%%%%%%%%%%%%%%%%%%%%%%%%%%%%%%%%%%%%%%%%
\subsection{Characterizing dynamic risk using deep learning}

To obtain a tractable optimization scheme, we restrict ourselves to policies represented by neural networks (the actor) whose parameters are denoted by $\theta$. Such parametric policies are denoted by $\pi^\theta$. 
Therefore, the approximate problem we tackle is
\begin{equation}
    \label{eq:approxprob}
    \theta^* = \underset{\theta}{\arg\!\min} \, \rho_{0,T}(c^{\pi^\theta}_0,\ldots, c^{\pi^\theta}_T ) = \underset{\theta}{\arg\!\min} \, V^{\pi^\theta}_0(s_0).
\end{equation}
The associated value functions (the critic) are also represented by a parametric function $V^\Psi$ with parameters $\Psi$. As such, for any policy $\pi^\theta$ we form some approximation $V^{\pi^\theta} \approx V^\Psi$ for some $\Psi$. Based on \eqref{eq:value function-nested}, this leads, for $t<T$, to the approximation
\begin{equation} \label{policyevaleq}
    V^\Psi_t(s_t) \approx \rho_t(c^{\pi^\theta}_{t+1} + V^\Psi_{t+1}(s_{t+1})). 
\end{equation}
The aim of the actor-critic algorithm is to progressively refine $\theta$ and $\Psi$ to reduce the objective function from \eqref{eq:approxprob} (the policy improvement step) and to have the relationship \eqref{policyevaleq} hold as closely as possible (the policy evaluation step).

For compatibility with the scoring function described, we follow \cite{CoacheJaimungalCartea2023} and separate the value function for dynamic CVaR into a VaR component $Z_t$ and a positive excess component (CVaR minus VaR) denoted by $A_t \geq 0$. This entails separating the critic parameter set into two: $\Psi=(\psi, \phi)$, where $\psi$ and $\phi$ are parameters of neural networks that approximate the VaR $Z$ and the excess $A$, respectively. For optimization stability, we use an ensemble of neural networks $\{Z^{\psi_t}_t, A^{\phi_t}_t\}_{t=0}^{T-1}$, with each time step having a pair of networks $(Z_t,A_t)$.  The set of time-indexed parameters is denoted by $\{\Psi_t\}^{T-1}_{t=0} = \{\psi_t, \phi_t\}^{T-1}_{t=0}$. For a fixed policy $\pi$ (we drop the notational dependence on the policy) we set
\begin{equation}
    V^{\Psi_t}_t(s_t) = Z^{\psi_t}_t(s_t) + A^{\phi_t}_t(s_t).
    \label{eq:valuesum}
\end{equation} 
Define optimized critic parameters $\{\Psi^*_t\}^{T-1}_{t=0}$ obtained through the policy evaluation steps
\begin{equation}
    \Psi^*_t = \underset{(\psi_t,\phi_t)}{\arg\! \min}  \, \mathbb{E} \left[ S(Z^{\psi_t}_t(s_t),V^{\Psi_t}_t(s_t), c_{t+1} + V_{t+1}^{\Psi^*_{t+1}}(s_{t+1}))\right],
\end{equation}
%\fredcomment{Do we need apostrophe in RHS for value function to explain this is target network.}
applied recursively for $t=T-1,\ldots,0$ with $V^{\Psi_{T}}_{T}\equiv 0$. We therefore have
\begin{align}
    Z^{\psi^*_t}_t(s_t) &\approx \mathrm{VaR}_{t,\alpha}(c_{t+1} + V^{\Psi^*_{t+1}}_{t+1}(s_{t+1}) ),
    \label{eq:Varinterm}
    \\ A^{\phi^*_t}_t(s_t) &\approx \mathrm{CVaR}_{t,\alpha}(c_{t+1} + V^{\Psi^*_{t+1}}_{t+1}(s_{t+1}) ) - \mathrm{VaR}_{t,\alpha}(c_{t+1} + V^{\Psi^*_{t+1}}_{t+1}(s_{t+1}) ).
    \label{eq:excessinterm}
\end{align}
%\fredcomment{same comment than above}

%where the expectation operates over the transition distribution $\mathbb{P}(s^{'}\mid s_t, a)\pi^{\theta}(a\mid s_t)$ which is policy-dependent.

%Given the condition that CVaR must be greater than the VaR for non-degenerate distributions, we approximate the value-function which represents the conditional CVaR of the continuation risk as a VaR component and strictly positive excess component. 

\subsection{Scoring function characterization \& geometry}
\label{sec:SGeo}
%Notice that Corollary~\ref{Cor:ElicitabilityCVaR} does not contain an additional real-valued function operating on the $\mathrm{VaR}$ estimate, since such a function can be some constant and leads to a simpler scoring function that directly emphasizes CVaR curvature. Thus, we do not consider characterizations of the score that utilize two real-valued non-constant functions, though this may prove to lead to more efficient optimization. 

\begin{figure}[h]
    \centering
    \includegraphics[width=\linewidth]{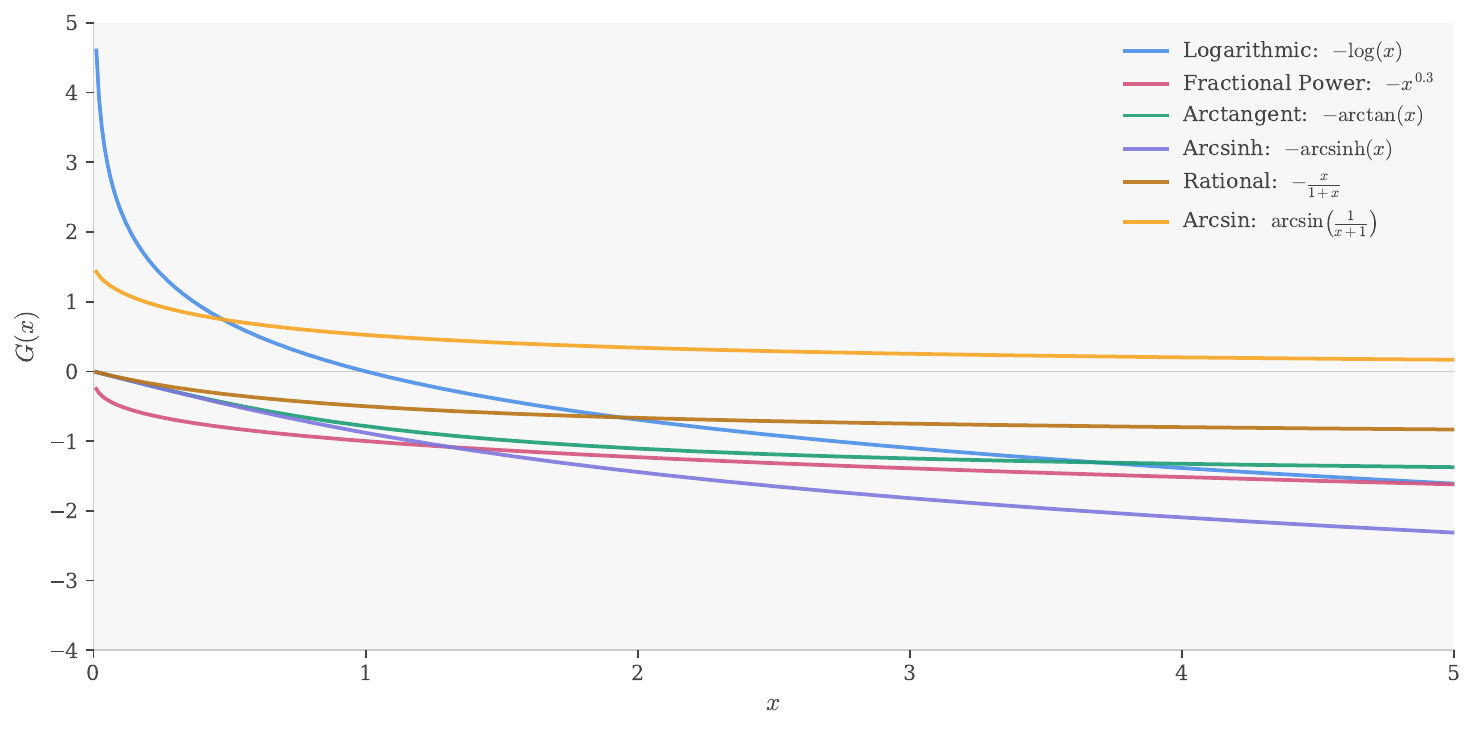}
    \caption{Comparison of the $6$ real-valued functions convex on the region $(0, \infty)$}
    \label{fig:functiongraphs}
\end{figure}

A central contribution of this paper lies in choosing varying characterizations of the scoring function, and additionally considering how to design the RL problem that leads to desirable optimization landscapes under such characterizations. First, the elimination of $H$ in the score requires that $G$ must be strictly convex and strictly decreasing for a strictly-consistent score.\footnote{For a strictly consistent scoring function, $G$ is strictly decreasing because its derivative is strictly negative, which follows from the fact that the mapping in \eqref{eq:Affinemap} must be strictly increasing for any $\mathfrak{a}_2$ when $H(z)=0$.} Under this guise, we can observe some appealing characteristics pertaining to the local curvature of our scoring function. We note that the Hessian of $\mathfrak{s}(\mathfrak{a}_1, \mathfrak{a}_2):=\mathbb{E}_{Y \sim F}\left[S(\mathfrak{a}_1, \mathfrak{a}_2, Y) \right]$ at the optimum $(\mathfrak{a}_1^*, \mathfrak{a}_2^*)$ where $f(x)$ is the Probability Density Function (PDF) of $Y$ is 
\begin{equation}
    \nabla^2 \mathfrak{s}(\mathfrak{a}_1^*, \mathfrak{a}_2^*) = 
    \begin{pmatrix} 
    \dfrac{ -f(\mathfrak{a}_1^*) G'(\mathfrak{a}_2^*)}{1-\alpha} & 0 \\[10pt] 
    0 & G''(\mathfrak{a}_2^*) 
    \end{pmatrix}.
    \label{eq:theHessian}
\end{equation}
See~\Cref{app:Hessian Derivation} for the derivation. We can observe that the Hessian evaluated at the optimum is positive definite if $G$ satisfies the conditions of a strictly-consistent score, and $f(\mathfrak{a}_1^*) > 0$, as the diagonal entries (and hence the eigenvalues) of the Hessian,
\begin{align*}
-\frac{f(\mathfrak{a}_1^*)G'(\mathfrak{a}_2^*)}{1-\alpha},
\qquad
G''(\mathfrak{a}_2^*)
\end{align*}
are strictly positive. Additionally, the zero off-diagonal entries indicate that the VaR and CVaR directions are locally decoupled at the optimum. In particular, a perturbation in the VaR direction does not affect the gradient in the CVaR direction, and a perturbation in the CVaR direction does not affect the gradient in the VaR direction. A novel contribution of our work is the examination of six possible characterizations of strictly consistent scoring functions arising from different choices of $G$ in our model-free environment. The choices of $G$ can be seen in~\Cref{tab:scoring-functions} and their graphs in~\Cref{fig:functiongraphs}. We choose $G$ to exhibit convexity on positive reals, with the offset $C$ term being added to $x$ for strict positivity. In this paper, we do not consider transformations in which $G$ rescales $x$ prior to the addition of $C$, which may lead to more desirable optimization regions for the scoring functions.
%; we instead use the shifted scoring function
% \begin{equation*}
% \begin{split}
% S(\mathfrak{a}_1, \mathfrak{a}_2, y) = 
%  - G(\mathfrak{a}_2+C) + G(y+C) + G'(\mathfrak{a}_2+C)\left[\mathfrak{a}_2 + \frac{1}{1-\alpha}\left(\mathfrak{a}_1\left(\Ind_{y > \mathfrak{a}_1} - (1-\alpha)\right) - y\Ind_{y > \mathfrak{a}_1}\right)\right].
% \end{split}
% \label{eq:CVaRScoreShift}
% \end{equation*}

\begin{figure}[ht]
    \centering
    \includegraphics[width=\linewidth]{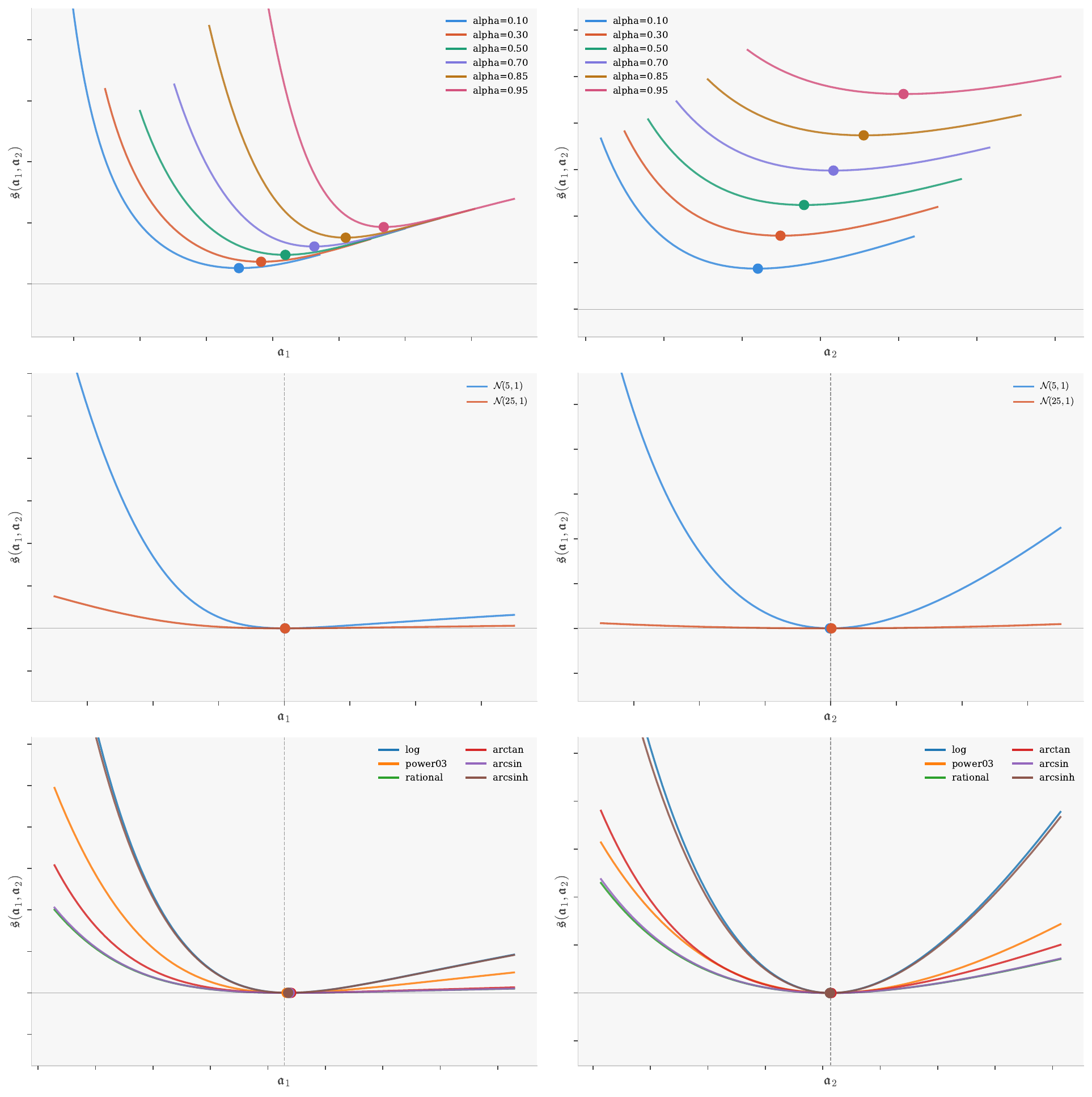}
    \caption{All datasets use $1,000,000$ samples. The top two plots compare curvature around the optimum $(\mathfrak{a}^*_1,\mathfrak{a}^*_2)$ for varying confidence levels $\alpha \in [10\%, 95\%]$, with $Y \sim \mathcal{N}(0,1)$ fixed, for the logarithmic scoring function $G(x) = -\log(x+C)$. The middle two plots compare curvature for $Y \sim \mathcal{N}(5,1)$ versus $Y \sim \mathcal{N}(25,1)$ at $\alpha = 95\%$, with the optimum shifted to a common origin for visual comparison. The bottom two plots compare curvature across scoring functions at $\alpha = 95\%$ on $Y \sim \mathcal{N}(0,1)$, with the shift constant $C$ set to the negative minimum of the sampled data so that all values are positive and the expected score evaluated at the optimum aligned at a common horizontal line to better visualize curvature.}
    \label{fig:scorelandscape}
\end{figure}

\begin{comment}
As an example, the scoring functions for the $G(x) = \log(x + C)$ and $G = -\mathrm{arcsinh}(x+C)$ choices (see the scoring functions for the remaining functions in~\Cref{app:remscores}) are
\begin{equation}
\begin{split}
S_{\text{log}} = \log \left(\frac{\mathfrak{a}_2 + C}{y+C} \right) - \frac{\mathfrak{a}_2}{\mathfrak{a}_2 + C} + \frac{\mathfrak{a}_1  (\Ind_{y \leq \mathfrak{a}_1} - \alpha) + y(\Ind_{y > \mathfrak{a}_1})}{(\mathfrak{a}_2 + C)(1-\alpha)} \quad \text{and}
\end{split}
\label{eq:logscore}
\end{equation}

\begin{equation}
    \begin{split}
        S_{\text{arcsinh}} = \operatorname{arcsinh}(\mathfrak{a}_2 + C) - \operatorname{arcsinh}(y + C) - \frac{
        \mathfrak{a}_2}{\sqrt{(\mathfrak{a}_2 + C)^2 + 1}} \\  +\frac{\mathfrak{a}_1  (\Ind_{y \leq \mathfrak{a}_1} - \alpha) + y(\Ind_{y > \mathfrak{a}_1})}{(\sqrt{(\mathfrak{a}_2 + C)^2 + 1})(1-\alpha)}.
    \end{split}
    \label{eq:hyperbolic_score}
\end{equation}
\end{comment}
\begin{table}[h]
\centering
\small
\setlength{\tabcolsep}{4pt}
\begin{tabular}{p{1.8cm} p{2.85cm} p{3.75cm} p{3.95cm}}
\toprule
\textbf{Class} 
& $G(x)$
& $G'(x)$
& $G''(x)$\\
\midrule

Logarithmic 
& $-\log(x+C)$ 
& $-\dfrac{1}{x+C}$ 
& $\dfrac{1}{(x+C)^2}$ \\[6pt]

Fractional-power
& $-(x+C)^{0.3}$ 
& $-0.3\,(x+C)^{-0.7}$ 
& $0.21\,(x+C)^{-1.7}$ \\[6pt]

Arctangent
& $-\arctan(x+C)$ 
& $\dfrac{-1}{1+(x+C)^2}$ 
& $\dfrac{2(x+C)}{(1+(x+C)^2)^2}$ \\[6pt]

Arcsinh
& $-\operatorname{arcsinh}(x+C)$ 
& $-\dfrac{1}{\sqrt{1+(x+C)^2}}$ 
& $\dfrac{x+C}{(1+(x+C)^2)^{3/2}}$ \\[6pt]

Rational
& $-\dfrac{x+C}{1+(x+C)}$ 
& $-\dfrac{1}{(1+x+C)^2}$ 
& $\dfrac{2}{(1+x+C)^3}$ \\[6pt]

Arcsin
& $\arcsin\!\left(\dfrac{1}{x+C+1}\right)$ 
& $\dfrac{-(x+C+1)^{-1}}
{\sqrt{(x+C+1)^2-1}}$ 
& $\dfrac{2(x+C+1)^2-1}
{(x+C+1)^2((x+C+1)^2-1)^{3/2}}$ \\

\bottomrule
\end{tabular}
\caption{Functions $G$ considered for inclusion in the scoring function \eqref{eq:CVaRScore2}, along with first and second derivatives w.r.t. $x$. The functions are valid for the domain $x \in(-C, \infty)$, where $C$ is a suitable positive real number.}
\label{tab:scoring-functions}
\end{table}
From our chosen $G$ functions, we may observe that both the first and second derivatives w.r.t. x approach 0 from below and above respectively for large $x$. This gives rise to a particular sensitivity property of the scoring function in~\Cref{eq:CVaRScore2}: as the CVaR optimum $\mathfrak{a}_2^*$ increases, curvature along both the VaR and CVaR axes decreases towards 0, and when the VaR optimum $\mathfrak{a}_1^{*}$ increases, curvature on the VaR scales with $f(\mathfrak{a}_1^*)$. These local properties can be visually examined through the global score landscape.~\Cref{fig:scorelandscape} aims to analyze how the convexity of function $s$, which is defined as the expectation of the scores of samples of the underlying distribution of $Y$, changes around the optimal point $(\mathfrak{a}^{*}_1,\mathfrak{a}^{*}_2)$ across different scoring function and varying distributional features. First, the top two plots analyze curvature for various CVaR confidence levels using the logarithmic scoring function.
%\fredcomment{Not sure 0-homogenous, since log(cx) is not equal to log(x).}
%\Shuyicomment{Clarified; the scoring function with $G(x) = -\log(x)$ is 0-homogeneous when $C=0$}
%\fredcomment{Even when C=0 it is not homogeneous... We need to relabel this property.}
As $\alpha$ increases, $\mathfrak{s}$ evaluated at the optimum rises since $\mathrm{VaR}_\alpha$ shifts further into the tail, while local convexity around the optimum stays roughly constant because the decreasing curvature from $G''(\mathfrak{a}_2^*)$ and growing $1/(1-\alpha)$ in~\Cref{eq:theHessian} result in weighting that roughly cancels out. The middle two plots compare curvature for $Y \sim \mathcal{N}(5,1)$ and $Y \sim \mathcal{N}(25,1)$ at a fixed $\alpha$, with the curves shifted so the minimum $\mathfrak{s}$ aligns for better comparison. One may remark that the $\mathcal{N}(5,1)$ score landscape is visibly more convex than $\mathcal{N}(25, 1)$. This is because the $C$ constant in $G$ must be set to a larger value for $\mathcal{N}(25,1)$ due to the rightward shift in the mean of the distribution, leading to an optimization problem that primarily operates in a region with flatter curvature. Lastly, the bottom two plots compare curvature for each scoring function around the optimum at a fixed $\alpha$ on $Y \sim \mathcal{N}(0,1)$, with the minimum $\mathfrak{s}$ aligned for better comparison. The log ($G(x) = -\log(x+C)$), arcsinh ($G(x) = -\text{arcsinh}(x+C)$), and power03 (fractional-power) ($G(x) = -(x+C)^{0.3}$) scoring functions (we refer to scoring functions based on the choice of $G$) have the steepest convexity near the optimum, while the arctan ($G(x) = -\text{arctan}(x+C)$), arcsin ($G(x) = \text{arcsin}(\frac{1}{1+x+C})$) and rational ($G(x) = -\frac{x+C}{x+C+1}$) scoring functions are visibly flatter. This follows from the fact that the log and arcsinh $G$ functions have second derivatives w.r.t. x that decay at the rate $x^{-2}$ for large $x$, whereas the arctan, arcsin, and rational $G$ functions have second derivatives w.r.t. x that decay at the faster rate $x^{-3}$, causing their curvature to approach zero more rapidly and resulting in flatter convexity. While the power03 $G$ function has a second derivative w.r.t. $x$ that decays at the rate $x^{-1.7}$, its coefficient of $0.21$ results in an overall lower magnitude that places it between the log and arcsinh $G$ functions and the arctan, arcsin, and rational $G$ functions. Thus, its curvature lies is relatively moderate as can be in the landscapes around the VaR and CVaR optimum. Along with these curvature properties, we may also observe that both the log and arcsinh $G$s have an unbounded range of $(-\infty, \infty)$, with first derivatives that decay roughly as $\frac{1}{x}$ which allow the $G$s to sustain unbounded growth (for a differentiable function to sustain unbounded growth while its rate of change tends to zero, that rate of change must decay sufficiently slowly that its integral remains unbounded). Meanwhile, the arctan, rational, and arcsin $G$s have first derivatives that decay according to $\frac{1}{x^2}$. The power03 function only has an unbounded range over the positive real line, with its derivative decaying as $0.3x^{-0.7}$. Again, despite this relative slower decay in comparison to the other $G$s, the coefficient $0.3$ results in a lower derivative magnitude overall. Taken together, the decay rates of $G'(x)$ and $G''(x)$ characterize how the corresponding scoring functions retain tail sensitivity as outcomes move into the tails. Based on these properties and the resulting asymptotic behavior of the scoring functions, we classify them into two categories. We refer to the log, power03, and arcsinh $G$ specifications as \emph{unbounded sublinear} scoring functions, since $G$ remains unbounded below, its first derivative vanishes asymptotically, and its magnitude grows more slowly than linearly. In contrast, we refer to the arctan, rational, and arcsin $G$ specifications as \emph{saturating} scoring functions, since their first derivatives vanish sufficiently rapidly for $G$ to approach a finite limit as its argument diverges. In~\Cref{sec:Expres}, this categorization allows us to clearly identify how the scoring function class influences the performance of the trained policies.
% To determine if these differences in the range, first derivatives and second derivatives of $G$ have an impact in the estimation of VaR and CVaR, we test these scoring functions on the same normal distribution used to generate the graphs of $s$ in the bottom two plots of~\Cref{fig:scorelandscape}. 

%%%%%%%%%%%%%%%%%%%%%%%%%%%%%%%%%%%%%%%%%%%%%%%%%%%%
\section{Model-free actor-critic reinforcement learning framework}
\label{sec:ACRL}
We follow the typical actor-critic RL framework, where some critic aims to gradually approximate the risk associated with the actor (policy), and the actor is improved through a policy gradient approach and minimizes the evaluated risk estimated by the critic. Importantly, the critic is only trained on freshly generated data from in-sample policies in an on-policy fashion. On-policy algorithms are less prone to state distribution mismatch and can lead to more stable gradient updates than off-policy methods, such as value-based Deep Q-Networks (DQN)~\citep{MnihEtAl2015} and Deep Deterministic Policy Gradient (DDPG)~\citep{LillicrapEtAl2015}, which typically reuse data generated by past policies through replay buffers. 
\subsection{Divided time horizon and sequential training}
\label{sec:SequentialShifting}
In traditional RL, value functions for all periods of a finite time horizon are learned simultaneously, making the training process increasingly challenging and potentially unstable as the time horizon and complexity of the environment increase. We propose a hybrid sequential training procedure that separates the time horizon into equivalent-sized portions and shifts the value (risk-to-go) function in order to ensure a stable sequential training procedure and guide the optimization problem to more desirable regions. In particular, because the value function may accumulate to large values due to the recursive structure of dynamic risk, we propose applying periodic downward shifts to the value function estimates. These shifts reduce the magnitude of the resulting value function and reshape the optimization landscape induced by the elicitability-based objective, making it more suitable for learning. Because dynamic CVaR is translation invariant, the optimal value function can be recovered after accounting for the applied shifts.

%\fredcomment{Why similarly efficient, why not more efficient (we are trying to beat the approach of simultaneous approx).}
%\fredcomment{Breaking down into subgroups allows shifting trick more efficiently (avoid accumulation of large values, lower convexity region).}

Shifting is applied separately (i.e. with different constants) and in sequence to the various sub-portions of the horizon, which leads to better use of high-convexity regions of functions $G$ and facilitates the optimization of the policy. We divide the time horizon $\mathcal{T}$ with $T+1$ time points into a set  $\mathcal{G} := \{0,....,\mathbb{G}\}$ of groups,
%\fredcomment{G is already used within the scoring function, we need another notation.}
where each group $g \in \mathcal{G}$ contains an integer number $d =T/(\mathbb{G}+1)$ periods
corresponding to times $\mathcal{T}_g :=\{gd,\ldots,{(g+1)d-1}\}$. When updating the value function at such time points,
we use target networks corresponding to value functions of periods $\bar{\mathcal{T}}_g :=\{gd+1,\ldots,(g+1)d\}$ periods (i.e. single-period offsets of $\mathcal{T}_g$). 
For the particular case of the last period of some group $g \not = \mathbb{G}$, the target network is updated with the \emph{frozen} value function of the first period of the subsequent group.

The training procedure starts from the last group $\mathbb{G}$, and trains the critic neural networks for a set number of gradient steps (each of which involves a number of target network updates). 
Then, we proceed recursively backward where on each iteration $g \in \mathcal{G} \textbackslash \mathbb{G}$, we freeze the networks for group $g+1$ (in fact what matters is freezing the value function at time $(g+1)d$, which is the first time point of group $g+1$) and update these corresponding to group $g$. This procedure is applied until all groups are ultimately updated, finishing with group $g=0$.
%\fredcomment{Are targets networks for non-terminal steps frozen during this, or they are updated? ANSWER: JUST SUBSEQUENT GROUP}

%The purpose of dividing the simultaneous value function estimation problem into a recursive estimation of multiple sub-problems in addition to the use of target networks is to further stabilize the training process by grounding the continuation risk by freezing the value function of certain periods to ensure that a stable training procedure in our environment.

% \label{sec:InputShifting}
%\Shuyicomment{This section is quite different from the others in that it proposes a way to "shift" the random variables, so that the critic learns the risk on a sequence of modified random variables (i.e., what the sequence of conditional dynamic risk is operating on) so that the random variables have lesser magnitude and whose expected score optimization landscapes operate with better curvature}

% A key benefit of our proposed framework is that we can periodically adjust the estimation problem to a well-defined optimization landscape.
As discussed above, the optimization of the score function $\mathfrak{s}(\cdot,\cdot)$ is easier when the convexity of $\mathfrak{s}$ is steep around the optimal point $(\mathfrak{a}_1^*, \mathfrak{a}_2^*)$. The optimization landscape can therefore be made more favorable by exploiting the translation property of the VaR and CVaR which entails shifting the value function by a deterministic constant,
%\fredcomment{Here we mention the value function being shifted}
which does not impact optimal actions and allows for full freedom to shift only based on optimization convenience considerations. Adjusting with different constants on different subgroups $g$ also allows adjusting for the accumulation of risk when iterating backwards through the time periods, which would have led to less than ideal curvature regions in absence of shifting corrections.

Consider a fixed policy $\pi^{\theta}$ and some deterministic values $b_0,\ldots,b_{\mathbb{G}}$. Define for the last group
\begin{align*}
    \tilde{V}_{T-1}(s_{T-1}) &:= \mathrm{CVaR}_{T-1, \alpha}(c_T + b_{\mathbb{G}}),
    \\ \tilde{V}_t(s_t) &:= \underbrace{\mathrm{VaR}_{t,\alpha}(c_{t+1} + \tilde{V}_{t+1}(s_{t+1}) )}_{ =: \tilde{Z}_t(s_t)} + \underbrace{\left(\mathrm{CVaR}_{t,\alpha}(c_{t+1} + \tilde{V}_{t+1}(s_{t+1}) ) - \mathrm{VaR}_{t,\alpha}(c_{t+1} + \tilde{V}_{t+1}(s_{t+1}) )\right) }_{  =: \tilde{A}_t(s_t)}
    \\  &= \left(Z_t(s_t) + b_{\mathbb{G}}\right) +  A_t(s_t), \quad t= \mathbb{G} d, \ldots,T-2,
\end{align*}
where the last equality is obtained through translation invariance. Furthermore, for prior groups $g< \mathbb{G}$ (i.e. times $t\in \mathcal{T}_g$) define,
\begin{align*}
    \tilde{V}_t(s_t) &:= Z_t(s_t) + \tilde{A}_t(s_t)  - (b_{g+1} - b_{g}) \text{ if } t= (g + 1)d - 1,
    \\ \tilde{V}_t(s_t) &:= \underbrace{\mathrm{VaR}_{t,\alpha}(c_{t+1} + \tilde{V}_{t+1}(s_{t+1}) )}_{ =: \tilde{Z}_t(s_t)} + \underbrace{\left(\mathrm{CVaR}_{t,\alpha}(c_{t+1} + \tilde{V}_{t+1}(s_{t+1}) ) - \mathrm{VaR}_{t,\alpha}(c_{t+1} + \tilde{V}_{t+1}(s_{t+1}) )\right) }_{  =: \tilde{A}_t(s_t)}
    \\  &= \left(Z_t(s_t) + b_g\right) +  A_t(s_t), \quad t= gd , \ldots, (g + 1)d - 2.
\end{align*}

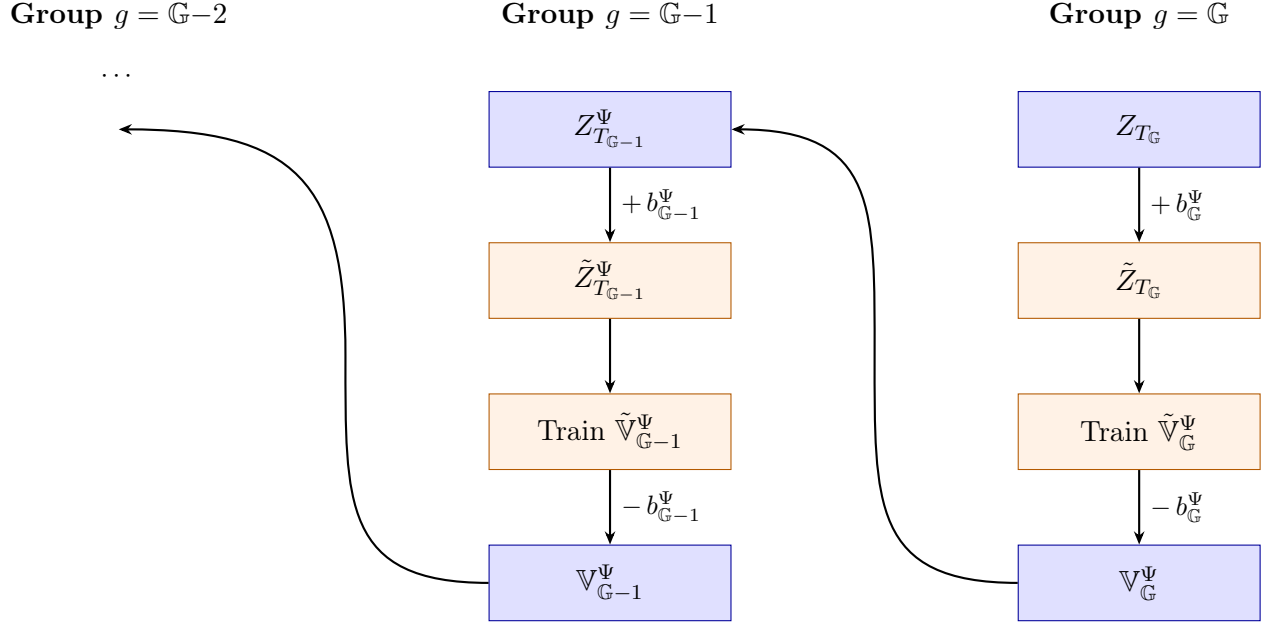
\begin{figure}[ht]
    \centering
    \begin{tikzpicture}[
        every node/.style={font=\small},
        rnode/.style={rectangle, draw=blue!60!black, fill=blue!12,
                      minimum width=3.2cm, minimum height=1.0cm, inner sep=4pt},
        cnode/.style={rectangle, draw=orange!70!black, fill=orange!10,
                      minimum width=3.2cm, minimum height=1.0cm, inner sep=4pt},
        arr/.style={-{Stealth[length=5pt,width=4pt]}, thick},
        lbl/.style={font=\footnotesize},
    ]

    %% ============================================================
    %% GROUP G  (right column)  x-centre = 12
    %% ============================================================
    \node[font=\bfseries\small] at (12, 5.5) {Group $g =  \mathbb{G}$};

    \node[rnode] (gZraw)   at (12,  4.0) {$Z_{T_ \mathbb{G}}$};
    \node[cnode] (gZshift) at (12,  2.0) {$\tilde{Z}_{T_ \mathbb{G}}$};
    \node[cnode] (gVtil)   at (12,  0.0) {Train $\tilde{\mathbb{V}}^{\Psi}_{ \mathbb{G}}$};
    \node[rnode] (gVrec)   at (12, -2.0) {$\mathbb{V}^{\Psi}_{ \mathbb{G}}$};

    \draw[arr] (gZraw)   -- node[right, lbl]{$+\,b^{\Psi}_ \mathbb{G}$}  (gZshift);
    \draw[arr] (gZshift) --                              (gVtil);
    \draw[arr] (gVtil)   -- node[right, lbl]{$-\,b^{\Psi}_ \mathbb{G}$}  (gVrec);

    %% ============================================================
    %% GROUP G-1  (left column)  x-centre = 5
    %% ============================================================
    \node[font=\bfseries\small] at (5, 5.5) {Group $g =  \mathbb{G}{-}1$};

    \node[rnode] (g1Zraw)   at (5,  4.0) {$Z^{\Psi}_{T_{ \mathbb{G}-1}}$};
    \node[cnode] (g1Zshift) at (5,  2.0) {$\tilde{Z}^{\Psi}_{T_{ \mathbb{G}-1}}$};
    \node[cnode] (g1Vtil)   at (5,  0.0) {Train $\tilde{\mathbb{V}}^{\Psi}_{ \mathbb{G}-1}$};
    \node[rnode] (g1Vrec)   at (5, -2.0) {$\mathbb{V}^{\Psi}_{ \mathbb{G}-1}$};

    \draw[arr] (g1Zraw)   -- node[right, lbl]{$+\,b^{\Psi}_{ \mathbb{G}-1}$}  (g1Zshift);
    \draw[arr] (g1Zshift) --                                   (g1Vtil);
    \draw[arr] (g1Vtil)   -- node[right, lbl]{$-\,b^{\Psi}_{ \mathbb{G}-1}$}  (g1Vrec);

    %% ============================================================
    %% INTER-GROUP ARROW: G -> G-1
    %% ============================================================
    \draw[arr]
        (gVrec.west)
        to[out=180, in=270, looseness=1.4]
        (8.5, 1.0)
        to[out=90, in=0, looseness=1.4]
        (g1Zraw.east);

    %% ============================================================
    %% CONTINUATION ARROW: G-1 -> G-2
    %% ============================================================
    \draw[arr]
        (g1Vrec.west)
        to[out=180, in=270, looseness=1.4]
        (1.5, 1.0)
        to[out=90, in=0, looseness=1.4]
        (-1.5, 4.0);

    \node[font=\bfseries\small] at (-1.5, 5.5) {Group $g =  \mathbb{G}{-}2$};
    \node[lbl]                  at (-1.5, 4.7) {$\cdots$};

    \end{tikzpicture}
    \caption{Illustration of the sequential training procedure. For each group $g \in \mathcal{G}$, define $\mathbb{V}_g := \{V^{\Psi_t}_t\}_{t=gd}^{(g+1)d-1}$. Define the endpoint of each group as $T_g = \max(\bar{\mathcal{T}}_g)$, with $T_{\mathbb{G}}=T$. The random variable at the terminal period $T_g$ is shifted up by $b^{\Psi}_g$, the value function $\tilde{V}^{\Psi}_{t}$ is trained on all time periods of the group $\mathcal{T}_g$ in the shifted space, and the recovered $V^{\Psi}_{t} = \tilde{V}^{\Psi}_{t} - b_g$ is frozen and passed to the preceding group $g{-}1$ to ground its terminal random variable. The process iterates backwards from $g = \mathbb{G}$ to $g = 0$.}
    \label{fig:bshift}
\end{figure}

As mentioned, because of the translation invariance property, optimal actions obtained when using the value function $\tilde{V}$ in the Bellman equation \eqref{eq:optimvalue function-nested} instead of $V$ are identical, and the optimal policy w.r.t. the modified value function $\tilde{V}$ is left unchanged. 
% Furthermore, a proper choice of constants $b_0,\ldots,b_{\mathbb{G}}$, values functions $\tilde{V}_t$ remain positive, which is desirable in our context since the functions $G$ (when $C=0$) possess desired properties for positive inputs. Thus, variables $b_0, \dots, b_{\mathbb{G}}$ are used for constant $\mathbb{G}$ within the specification of $G$. 
In practice, $b_g$ for a given group $g$ can be calculated as the negative sample minimum of the target random variable at ${(g+1)d-1}$ (i.e. the last time period of the group $g$) of some batch, ensuring that the shifted terminal target is always positive. As one may note, the shift applied to the terminal target random variable only propagates to the first target random variable after a sufficient number of training iterations and consequently does not guarantee that all target random variables are positive at the beginning of training. The $C$ term in the $G$ functions acts as the necessary additive shift (which is kept constant across all groups) to ensure all target random variables are initially positive. Although $b^{\Psi}_g$ will change as new batches are generated by the policy, the sampling variability can be mitigated using a sufficiently large batch size. %As the algorithm learns a stable policy, the policy-induced changes in $b_g$.
\Cref{fig:bshift} shows the training procedure visually, highlighting connections between original and shifted value function components during the training.

\subsection{Actor-critic RL algorithm}

Given a batch of $B$ $(s^{(b)}_t, a^{(b)}_t, s^{(b)}_{t+1}, c^{(b)}_{t+1})$ transitions for $t \in \mathcal{T}_g$ corresponding to a certain group $g\in \mathcal{G}$ and $b \in \mathcal{B}$ where $\mathcal{B} := \{1, \dots, B\}$,
the critic loss for some non-terminal group $g$ with starting and ending time period $t_{beg}$ and $t_{end}$ %and $d$ time periods 
%\fredcomment{Shall we write $t_b-t_a$ instead of $d$?}
can be defined as
\begin{equation}
\begin{split}
\mathcal{L}^{g}_{\Psi} = \frac{1}{d \times B}\sum_{t=t_{beg}}^{t_{end}-1}\sum_{b=1}^{B} S(Z^{\psi_t}_{t}(s_t; \pi^{\theta}), V^{\Psi_t}_{t}(s_t; \pi^{\theta}), c_{t+1} + V^{\Psi^{'}_{t+1}}_{t+1}(s_{t+1}; \pi^{\theta})) + \\
\frac{1}{d \times B}\sum_{b=1}^{B} S(Z_{t_{end}}^{\psi_{t_{end}}}(s_t; \pi^{\theta}), V_{t_{end}}^{\psi_{t_{end}}}(s_t; \pi^{\theta}), c_{t+1} + V_{t_{end}+1}^{\Psi^{'}_{t_{end}+1}}(s_{t_{end}+1}; \pi^{\theta}) - b^{\Psi}_{g+1} + b^{\Psi}_{g}).
\end{split}
\label{eq:criticlossinterm}
\end{equation}
 Note that the value of the next state is calculated using the target network, whose parameters are denoted by $\Psi^{'}$. We use hard target updates where the parameters are copied from the critic's parameters and then held frozen for a fixed number of updates before being periodically updated again, thereby stabilizing the random variables on which the dynamic risk is calculated.
%\fredcomment{Why a difference in b's? Why do b's depend on psi, is it because they are the essential infimum? How are b's decided in practice?}
%\fredcomment{Need to define notation $g(t_b)$}
Furthermore, the critic loss for the terminal group with starting time period $t_{beg}$ can be defined as
\begin{equation}
\begin{split}
\mathcal{L}^{\mathbb{G}}_{\Psi} = \frac{1}{d \times B}\sum_{t=t_{beg}}^{T-1}\sum_{b=1}^{B} S(Z^{\psi_t}_{t}(s_t; \pi^{\theta}), V^{\Psi_t}_{t}(s_t; \pi^{\theta}), c_{t+1} + V^{\Psi^{'}_{t+1}}_{t+1}(s_{t+1}; \pi^{\theta}) + b^{\Psi}_{G} \mathds{1}_{ \{t+1=T\}  }))  %+\\ \frac{1}{d \times B}\sum_{b=1}^{B}S(Z^{\psi_{T}}_{T}(s_{T}; \pi^{\theta}), V^{\Psi_{T}}_{T}(s_{T}; \pi^{\theta}), c_{T} + b^{\Psi}_G).
\end{split}
\label{eq:criticlossterm}
\end{equation}
% \fredcomment{Index of $\mathcal{L}$ is different in 3.8 and 3.9. Make homogenous (and dependent on group g).}  
% \fredcomment{What do we mean by the target network? Do you mean 'prime' instead of 'bar'?}
We follow the result of Theorem~6.3 of~\citet{CoacheJaimungalCartea2023} to obtain the gradient of the value function $\nabla_{\theta}V(s; \pi^\theta)$ in the context of stochastic policies which we are using in this work. Given a batch of $B$ transitions $(s^{(b)}_t, a^{(b)}_t, s^{(b)}_{t+1}, c^{(b)}_{t+1}, \log (\pi^{\theta}(a^{^{(b)}}_t \mid s^{(b)}_t))$ for $t \in \mathcal{T}$ and $b \in \mathcal{B}$,
the actor loss that approximates this gradient is
\begin{equation}
%\begin{split}
\mathcal{L}^{\theta}_{\text{actor}} 
= \ \frac{1}{(1-\alpha)T B}\sum_{t=0}^{T -1}\sum_{b=1}^{B}
\Big[\big(c_{t+1} + V^{\Psi_{t+1}}_{t+1}(s_{t+1};\pi^{\theta}) - Z^{\psi_t}_{t}(s_t;\pi^{\theta}) - b^{\Psi}_{\upsilon(t)}\big)_{+}
\nabla_{\theta}\log\pi^{\theta}(a\mid s_t)\big|_{a=a_t}\Big] 
%\\+ &\ \frac{1}{(1-\alpha) T B}\sum_{b=1}^{B} \Big[\big(c_T - Z^{\psi_T}_{T}(s_T;\pi^{\theta}) -
%b^{\Psi}_{G}\big)_{+}
%\nabla_{\theta}\log\pi^{\theta}(a\mid s_T)\big|_{a=a_T}\Big],
%\end{split}
\label{eq:actorloss}
\end{equation}
where $\upsilon(t) = \lfloor t / d \rfloor$.
It is important to note that because the critic is frozen during the actor update, a common choice in actor-critic methods, the loss does not include the additional terms arising from the critic's own dependence on $\theta$.\footnote{There are solutions proposed in the literature \citep[see for instance][]{HuangEtAl2021} to take into consideration the fact that the policy is not left untouched during training iterations, but we do not pursue such approach.}  %the literature one can either estimate the transition probabilities to recover the full correct gradient (see ~\citet{HuangEtAl2021}) and re-weight the loss or add the correction term.
 %\fredcomment{The term refers to policy update}
 %However, an actor-critic algorithm that maintains relatively conservative actor updates between critic updates can remedy this issue by ensuring that the degree of state distortion is limited. 
 %\fredcomment{This in unclear to me.}
 
The full algorithm is outlined in~\Cref{alg:grouped_value_update}, with hyperparameter values provided in~\Cref{app:hypACAO}. To address the sparse gradient contributions to the actor loss at high confidence levels, the algorithm weights the actor mini-batch size $B_2$ by $1/(1-\alpha)$, ensuring that a sufficient number of transitions contribute to the actor update. 
%\fredcomment{Is this done for all confidence levels?}
%\Shuyicomment{Yes, the same algorithm is used for varying $\alpha$}
%\Shuyicomment{Added the explanation of scaled batch size}
\begin{algorithm}[t]
\caption{Actor-critic Algorithm}
\label{alg:grouped_value_update}
\begin{algorithmic}

\Require Set of group periods $\mathcal{G} = \{0,\ldots,\mathbb{G}\}$ dividing the time horizon into equivalently sized sequences of one-step periods with each group containing $d$ periods each defined as $\mathcal{T}_g = \{t_{beg},\ldots,t_{end}\}$ with $t_{beg}$ and $t_{end}$ denoting the first and last time periods respectively of some group $g$, epochs $E$, critic iterations per group $E_1$, actor iterations $E_2$, batch size $B$, critic minibatch size $B_1$, actor minibatch size $B_2$, batch index sets
$\mathcal{B} = \{1,\dots,B\}$, $\mathcal{B}_1 = \{1,\dots,B_1\}$, and $\mathcal{B}_2 = \{1,\dots,\frac{B_2}{1-\alpha}\}$, target update frequency $M$, learning rates of $\eta^{\Psi}_{g}$ and $\eta^{\theta}$ for the critics of each group and actor respectively, critic neural networks as $\{\Psi_t, \Psi'_t\}_{t=0}^{T-1}$ with parameter set $\Psi = \{\psi, \phi\}$ for $Z^{\psi_t}_t, A^{\psi_t}_t$ and $\Psi'_t$ as the parameter sets for the target neural networks, and actor $\pi^\theta$ with parameter set $\theta$
\For{$e = 1, \dots, E$} 
\State Generate $B$ trajectories under $\pi^\theta$
    \For{$g = \mathbb{G}, \dots, 0$}
        \State Slice batch paths into one-step transitions $\{(s^{(b)}_t, a^{(b)}_t, c^{(b)}_{t+1}, s^{(b)}_{t+1}) \mid t \in \mathcal{T}_g,\ b \in \mathcal{B}\}$
        \For{$e_1 = 1, \dots, E_1$}
            %\State Reset the gradients w.r.t.\ $\{\Psi_t\}_{t_a}^{t_b}$ to zero
            \State Sample minibatch of size $B_1$
            \State Compute critic loss and associated gradient
            \State Update $\{\Psi_t\}_{t \in \mathcal{T}_g}$ via Nesterov-accelerated SGD
            \If{$e_1 \bmod M = 0$}
                \State $\Psi'_{t+1} \gets \Psi_{t+1}$ for $t \in \mathcal{T}_g \setminus \{t_b\}$ \Comment{Hard update target networks}
                \If{$g \neq \mathbb{G}$}
                    \State $\Psi'_{t_{end}+1} \gets \Psi_{t_{end}+1}$
                \EndIf
            \EndIf
            \State Decay $\eta^{\Psi}_{g}$
        \EndFor
        \State Freeze $\{\Psi_t\}_{t \in \mathcal{T}_g}$
    \EndFor
    \For{$e_2 = 1, \dots, E_2$}
        %\State Zero out the gradients w.r.t.\ $\theta$
        \State Generate $B_2/(1-\alpha)$ trajectories under $\pi^\theta$
        % \State Sample minibatch of size $\{(s^{(b)}_t, a^{(b)}_t, c^{(b)}_t, s^{(b)}_{t+1}) \mid t \in \mathcal{T},\ b \in \frac{\mathcal{B}_2}{1-\alpha}\}$
        \State Compute actor loss and associated gradient \Comment{Use all trajectories per actor step}
        \State Update $\theta$ via Adam 
        \State Decay $\eta_\theta$
    \EndFor
\EndFor
\State \Return $\{\Psi_t\}_{t=0}^{T-1}$, $\pi^\theta$
\end{algorithmic}
\end{algorithm}

%\fredcomment{What is the current epoch? Do you loop through all groups within a single epoch? This needs clarification. ANSWER: epoch is one training iteration for the actor, and then critic.}

%%%%%%%%%%%%%%%%%%%%%%%%%%%%%%%%%%%%%%%%%%%%%%%%%%%%%%%
%%%%%%%%%%%%%%%%%%%%%%%%%%%%%%%%%%%%%%%%%%%%%%%%%%%%%%%

\section{Methodology}
\label{sec:ExpMeth}
\subsection{Experimental setup}
In our experiment, we apply the actor-critic algorithm in~\Cref{alg:grouped_value_update} to the dynamic hedging problem described below across the $6$ strictly-consistent scoring functions mentioned in~\Cref{sec:SGeo} and four confidence levels of $\alpha \in \{92.5\%, 95\%, 97.5\%, 99\%\}$ for both a dynamic and static risk objective. The use of high risk thresholds allows us to demonstrate that our dynamic risk algorithm can learn policies that effectively target the tail of the remaining-cost distribution. At high thresholds, these policies increasingly approach the worst-case objective of minimizing maximum remaining costs. We fix the global random seed and initialize the neural networks consistently across runs for reproducibility. Randomized components, including minibatch selection and path generation, are controlled through seeded random-number generators. For path generation, the seed is determined by the training iteration and phase, ensuring that corresponding critic and actor updates are exposed to the same samples across runs. Diagnostic paths are generated using the same principle during evaluation. We also use large batch sizes throughout training to reduce sampling variability in the critic and actor losses and their gradients. The code to run the actor-critic RL algorithm can be found in the \href{https://github.com/shuyizhang01/Time-consistent_Deep-hedging}{GitHub repository}.

\subsection{Hedging environment}
The financial problem we study consists of dynamically hedging a European basket call option with a $T=252$ day (one-year) maturity and strike price of $K=100$ in a time-consistent fashion. The basket option payoff is calculated as the weighted arithmetic sum of $N=4$ bank assets indexed $i=1,\dots,N$: JPMorgan Chase \& Co. (JPM), Bank of America Corporation (BAC), Wells Fargo \& Company (WFC), and Citigroup Inc. (C), where the vector of time $t$ stock prices is denoted by $S_t = \{S_{i,t}\}_{i=1}^{N}$. All initial asset prices are set to $100$ dollars, and the basket weights we use are $\{w_i\}_{i=1}^{N} = (0.0775, 0.4434, 0.2649, 0.2142)$, corresponding to JPM, BAC, WFC, and C respectively. To replicate realistic financial market dynamics, we model the evolution of bank stock prices over the hedging horizon using a scalable DCC-GARCH(1, 1) framework with a Student-$t$ copula and Gaussian marginals, explicitly capturing stochastic volatility, dynamic correlations and tail dependence (see \cref{app:Env}). The model parameters are calibrated using ten years of historical daily returns for the four underlying bank stocks, from December 23, 2015, to December 20, 2025, to reflect a broad range of historical market conditions. Details of the calibration procedure and final parameter values are provided in \cref{app:Env}. For the hedging scenario, we assume that a hedging agent writes (sells) a basket call option for a premium $\mathcal{V}_0$. On each daily period $[t,t+1)$, the action vector $a_t=(a_{t,1},\ldots,a_{t,4}) \in \left[-2, 2\right]^4$ determines the number of shares of the four underlying assets they hold to hedge the risk of the option. When the basket option expires, the trader pays back the terminal value of the basket $\mathcal{V}_T$ and liquidates all positions. The cost $c_{t+1}=-(W_{t+1} - W_t)$ incurred on each period is defined as minus the change in the trader's net wealth whose dynamics is defined below.
We assume that the trader has access to a risk-free instrument where cash investments grow at a constant daily risk-free rate of $\bar{r}_f=0.0087\%$ which corresponds to an annualized rate of $2.11\%$. This constant rate is calculated as the average daily continuously compounded risk-free rate over the ten-year historical period based on 3-month U.S. Treasury bill yields. A proportional transaction cost rate of $\kappa = 0.1\%$ is applied to trades. Trading is assumed not to affect the market.

%\rc{A key note in our cost definition is that we also subtract the corresponding derivative price from the wealth prior to the terminal period to ensure that the terminal cost is similar in magnitude to other costs; because of this adjustment, we must also subtract the corresponding derivative price to all prior wealth amounts to ensure that the summation of costs remains a viable terminal hedging error objective.}
%\fredcomment{I am not sure I get the red part above, does it simply mean we report the wealth net of the derivative's price? If so we can maybe remove and simply call it net wealth.}
Where the basket option price $\mathcal{V}_{t}$ is estimated by the surrogate neural network described in~\Cref{sec:NNP}, the evolution of wealth for any $t \in \mathcal{T} \setminus \{\mathcal{T}\}$ is given by
\begin{equation} \label{eq:hedging}
    W_{t+1} = (W_t -a_{t} \cdot S_t - \kappa S_t \cdot |a_{t} - a_{t-1}|) e^{\bar{r}_f} +a_{t} \cdot S_{t+1} - \mathcal{V}_{t+1},
\end{equation}
% \begin{subequations}\label{eq:hedging}
% \begin{alignat}{2}
%   C_0      &= \mathcal{V}_0,
%     &\quad w_0      &= C_0; \\
%   C_t^{*}  &= C_t - (a_t - a_{t-1}) \cdot S_t - \lambda \, S_t \, |a_t - a_{t-1}|,
%     &\quad w_t      &= C_t + a_{t-1} \cdot S_t  - \mathcal{V}_{t}; \label{eq:postcash}\\
%   C_{t+1}  &= e^{r\Delta t} C_t^{*},
%     &\quad w_{t+1}  &= C_{t+1} + a_t \cdot S_{t+1} - \mathcal{V}_{t+1}; \\
%   C_T      &= e^{r\Delta t} C_{T-1}^{*} + a_{T-1} \cdot S_T - \lambda \, S_T \, |a_{T-1}| - \mathcal{V}_T,
%     &\quad w_T      &= C_T; \quad  \text{and} \\
% %  c_t      &= -(w_{t+1} - w_t).
% \end{alignat}
% \end{subequations}
where $W_0 = \mathcal{V}_0$\footnote{The transaction cost associated with liquidation $\kappa S_T \cdot{a_{T-1}}$ is additionally subtracted from $W_{T-1}$ to calculate $W_T$.}.
Thus, we define the cash account value as
\begin{equation*}
    \mathbb{C}_{t+1}  = e^{\bar{r}_f} \left( \mathbb{C}_t - (a_t - a_{t-1}) \cdot S_t - \kappa \, S_t \cdot |a_t - a_{t-1}|\right).
\end{equation*}
with $\mathbb{C}_0=\mathcal{V}_0$. 
%\fredcomment{Could we summarize that as $w_{t+1} = (w_t -a_{t+1} S_t - (a_{t+1} - a_{t})\lambda S_t) e^{r \Delta_t} +a_{t+1} S_{t+1} - \mathcal{V}_{t+1}$ with $w_0 = C_0 = \mathcal{V}_0$?}
%\fredcomment{Right now 4.1b and 4.1d give two definitions of wealth in contradiction.}
%\Shuyicomment{I fixed 4.1b to use the precash instead of post-cash which is now consistent with my experiment and removes the contradiction.}
%\fredcomment{Right equation (4.1b), should it be $a_t$ instead of $a_{t-1}$?}

%\fredcomment{Lowercase $s$ is used both as the state and the expected scoring function. Maybe we should adjust that. }
%\Shuyicomment{I}

To ensure that the hedging agent is able to perform across a relatively diverse set of paths, we randomize initial asset prices around their initial value and each asset's initial variance around the respective unconditional long-run value, with correlations starting at their long-run unconditional level. This is done by sampling from a centered normal distribution with a small standard deviation and applying a logarithmic shock to the initial stock price and variances.
All initial values can be seen in~\Cref{app:Env}. 

For numerical convenience, we normalize some of the inputs from the state space. At time $t$, given $h_t = \{h_{i,t}\}_{i=1}^{N}$ and $\bar{h} = \{\bar{h}_{i}\}_{i=1}^{N}$, the state is defined as $s_t = (S_t / K, h_t / \bar{h}, a_{t-1}, \operatorname{vec}_{\triangle}(R_t), \frac{T-t}{T})$ 
where $S_t / K$ represents the assets' moneyness values (underlying prices divided by the strike), $h_t / \bar{h}$ represents the assets' conditional variances normalized by their corresponding unconditional variances (which is calculated according to the Heston-Nandi GARCH parameters described in~\Cref{app:Env}), $a_{t-1}$ represents the previous actions of the hedging agent, $
\operatorname{vec}_{\triangle}(R_t)
=
(R_{t,ij})_{1 \leq i < j \leq n}
$, represents the pairwise conditional correlations, and
% The normalized cash account is defined as a sigmoid transformation of the ratio between the cash position and the current portfolio notional exposure:
% \begin{equation}
% \bar{C}_t
% =
% \sigma\!\left(
% \frac{C_t}{\left|\sum_{i=1}^{d} a_{t-1}^{(i)} S_t^{(i)}\right|}
% \right).
% \end{equation}
% The sigmoid function $\sigma(x) = (1 + e^{-x})^{-1}$ ensures the feature is bounded in $(0,1)$.
$(T  - t) / T$ represents the remaining time to maturity in years. We use the pairwise conditional correlations from $R_t$ rather than the diagonal and upper-triangular elements of the DCC proxy matrix $Q_t$ as it yields a more parsimonious representation of the evolving dependence structure while retaining the information relevant for hedging.
 The cash account is not included in the state as the CVaR risk measure is translation invariant.
% Although the cash amount is theoretically unnecessary to include since $\mathrm{CVaR}$ is a translation invariant risk measure, we include it anyways as our hedging problem is not transactionless and we also speculate that it may help in training our actor.
%\fredcomment{Have you checked this assumption? We would definitely have a question from refs about this.}
%\rc{Note that the inclusion of the cash-account does not change the necessary condition that the transition probability is only dependent on $\theta$ through the probability distribution of the policy for the actor loss in~\Cref{eq:actorloss}, since $B_t$ is not dependent on $a_t$ see~\Cref{eq:postcash}.}

\subsection{Benchmarking with a static risk measure}
As a benchmark, we train a static risk agent using a Monte Carlo policy gradient approach. Such approach involves directly differentiating through the terminal loss, which is reminiscent of that presented in \cite{Buehler2019}. Whereas the dynamic risk model solves \eqref{eq:approxprob} based on \eqref{nestRMdef} and $\rho_t= \text{CVaR}_{t,\alpha}$, the static risk model uses the \cite{RockafellarUryasev2000} CVaR representation and minimizes
\begin{equation}
\mathrm{CVaR}_\alpha
\left(
\sum_{t=1}^{T} c_t
\right)
=
\min_{q \in \mathbb{R}}
\left\{
q
+
\frac{1}{B(1-\alpha)}
\sum_{b=1}^{B}
\left[
\left(
\sum_{t=1}^{T} c_t
\right)
-q
\right]_+
\right\}
\label{eq:staticactorloss}
\end{equation}
where $q$ is a smooth auxiliary variable representing the $\mathrm{VaR}_{\alpha}$ of the summed costs and is jointly optimized with the actor neural network parameters during training. In a differentiable environment, the training procedure remains the same as~\Cref{alg:grouped_value_update}, excluding the critic training and replacing~\Cref{eq:actorloss} with~\Cref{eq:staticactorloss}. Notably, \citet{godin2016minimizing} showed that the static risk objective that optimizes $\mathrm{CVaR}_{\alpha}$ over terminal costs is time-consistent w.r.t. a conditional $\mathrm{CVaR}_{\alpha}$ with a varying $\alpha$ over time, up to technical conditions. However, since it is non-trivial to derive the varying $\alpha$ sequence, we primarily focus our analysis through the lens of a precommitment (static) vs. time-consistent (dynamic) hedging strategy and explore the differences between the two.
%the static risk objective which implies a varying dynamic risk measure that is equivalent in magnitude to the dynamic risk of the dynamic risk objective, 
%we cannot directly compare the time-consistent formulations of both approaches. Thus, 

\subsection{Policy parameterization}
The actor under both dynamic and static risk is characterized as a simple feed-forward neural network, both outputting allocations to each asset. For the dynamic risk policy, the actor learns a normal distribution $\mathcal{N}$ where $\pi^{\theta}(\cdot \mid s) := \mathcal{N}(o_{\theta}(s), \sigma^2 I)$ over the action space with $I$ being the identity matrix and $o_{\theta}(s)$ is the mean vector produced by the network. Applying the reparameterization trick allows for efficient gradient computation by sampling actions through $o_\theta(s) + \sigma \varepsilon$ where $\varepsilon \sim \mathcal{N}(0, I)$. The static risk actor is fully deterministic and outputs actions within the same range as the dynamic risk actor. Specifics of the algorithm, including the feed-forward neural network architectures and associated hyperparameters, are provided in~\Cref{app:hyp}.

%\fredcomment{Why do we want that if we check performance on a single set of starting value (do we also perturb in test set)?}

%\Shuyicomment{Yes, we also perturb in the test set for the important graphs/tables; for instance, the P\&L dist uses a perturbed test set but the portfolio delta curve does not otherwise the shape would be too spread out}

%%%%%%%%%%%%%%%%%%%%%%%%%%%%%%%%%%%%%%%%%%%%%%%%%%
\section{Experimental Results}
\label{sec:Expres}
In the following section, we present results on the hedging policies in our environment from the training procedure, which we outline below.

\subsection{Hedging performance}
\begin{figure}[h]
    \centering
    \includegraphics[width=\linewidth]{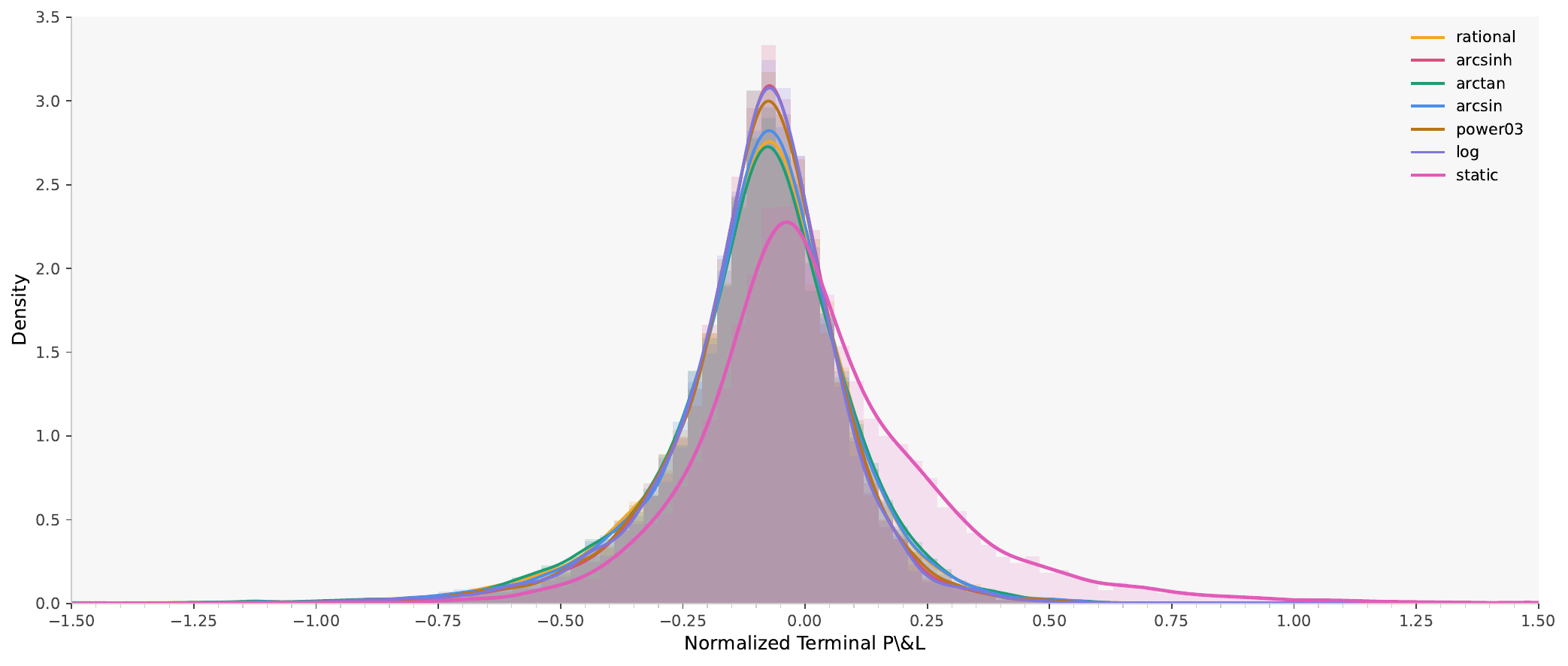}
    \caption{Comparison of normalized terminal P\&L (P\&L divided
    by initial option price for each path) densities for the dynamic and static risk models at the $\alpha = 95\%$ confidence level across the $6$ scoring functions $G$ for
    the dynamic risk model and for the static risk model tested across the $10,\!000$-path test set with initial state perturbation. With fixed initial conditions, the basket surrogate prices $\mathcal{V}_0$ at $12.04$ dollars.}
    \label{fig:pnldist}
\end{figure}

To evaluate the hedging policies under both the dynamic and static risk frameworks, we use test sets of $10,\!000$ paths with either perturbed or fixed initial conditions, where the former follows the same initial distribution as the training set and the latter uses unperturbed initial stock prices and unconditional long-run variance and pairwise correlations. The specific test set used for each figure and table is indicated where relevant. The dynamic risk policy is executed deterministically by taking the mean action $o_\theta(s)$ of the learned policy distribution when evaluating the hedging performance. We additionally consider the nested approach from \cite{CoacheJaimungal2024} to validate the accuracy of the RL-based critic. %\rc{Such an approach involves training a critic stochastically, since the value function estimates dynamic risk of a stochastic policy during training.}
%\fredcomment{I am not sure I get the above sentence; don't all approach train a critic stochastically? Maybe I am missing something.}
%\Shuyicomment{Yes, I added this so readers didn't think that I trained the nested critic with a deterministic actor (mentioned in the prev sentence), but now that I read it back we could remove it.}
%with the same initial distribution as the training set, and calculate the terminal hedging error for each trajectory.
%See~\Cref{fig:pnldist} for a comparison of the dynamic and static risk hedging policies. First, as our aim for both dynamic and static  risk models is to effectively mitigate the risk of the European basket option up to its maturity, we utilize the terminal P\&L (equivalent to the final portfolio value and hedging error) to measure hedging performance.
In~\Cref{tab:HedgeStats}, we report descriptive statistics (mean, standard deviation, and CVaR at various confidence levels) for the negative terminal P\&L distributions under both dynamic and static risk objectives. Throughout this section, models identified by a scoring function denote dynamic risk models, while models labeled as static denote static risk models. The static CVaR metrics are calculated as the upper tail of the negative P\&L. All other metrics are calculated on the non-negated P\&L. \Cref{fig:pnldist} shows the respective distributions for all dynamic and static risk models across only a normalized P\&L, where the normalization is performed by dividing each test set path's terminal P\&L by the initial basket option price. 
\begin{table}[htbp]
\centering
\small
\begin{tabular}{llrrrrrr}
\toprule
Confidence level & Model & Mean & Std & $\mathrm{CVaR}_{92.5\%}$ & $\mathrm{CVaR}_{95\%}$ & $\mathrm{CVaR}_{97.5\%}$ & $\mathrm{CVaR}_{99\%}$ \\
\midrule
\multirow{7}{*}{$\alpha=92.5\%$}
 & log      & -1.1197 & 1.7497 & 4.5702 & 4.9721 & 5.6579 & 6.5719 \\
 & power03  & -1.1933 & 1.7290 & 4.5799 & 4.9816 & 5.6587 & 6.5686 \\
 & arcsinh  & -1.1297 & 1.7521 & 4.5714 & 4.9751 & 5.6788 & 6.6379 \\
 & arctan   & -0.9786 & 1.6969 & 4.4233 & 4.8568 & 5.5935 & 6.5026 \\
 & rational & -0.9570 & 1.6498 & 4.3107 & 4.7327 & 5.4522 & 6.3507 \\
 & arcsin   & -0.9704 & 1.6863 & 4.3989 & 4.8347 & 5.5721 & 6.4766 \\
 & \emph{static} & 0.2936 & 2.5728 & 3.7648 & 4.0910 & 4.6219 & 5.2902 \\
\midrule
\multirow{7}{*}{$\alpha=95\%$}
 & log      & -1.1143 & 1.7095 & 4.4911 & 4.8922 & 5.5790 & 6.4954 \\
 & power03  & -1.0899 & 1.7319 & 4.4520 & 4.8351 & 5.4579 & 6.3675 \\
 & arcsinh  & -1.0901 & 1.7050 & 4.4531 & 4.8397 & 5.5148 & 6.4274 \\
 & arctan   & -1.1250 & 1.9440 & 4.9344 & 5.3720 & 6.1150 & 7.0660 \\
 & rational & -1.1227 & 1.9015 & 4.8756 & 5.3220 & 6.0702 & 6.9918 \\
 & arcsin   & -1.0913 & 1.8746 & 4.7818 & 5.2199 & 5.9494 & 6.8820 \\
 & \emph{static} & 0.2161 & 2.6423 & 3.8546 & 4.1595 & 4.6521 & 5.2612 \\
\midrule
\multirow{7}{*}{$\alpha=97.5\%$}
 & log      & -1.3074 & 2.1361 & 5.4259 & 5.8652 & 6.5945 & 7.6098 \\
 & power03  & -1.2588 & 2.0449 & 5.1843 & 5.5995 & 6.2991 & 7.2472 \\
 & arcsinh  & -1.2600 & 2.1102 & 5.3637 & 5.8009 & 6.5641 & 7.5813 \\
 & arctan   & -1.3347 & 2.3859 & 5.7730 & 6.2027 & 6.9023 & 7.7148 \\
 & rational & -1.3177 & 2.3300 & 5.6249 & 6.0475 & 6.7539 & 7.5905 \\
 & arcsin   & -1.3213 & 2.3513 & 5.6528 & 6.0803 & 6.7923 & 7.6186 \\
 & \emph{static} & 0.1443 & 2.7486 & 4.0504 & 4.3420 & 4.8096 & 5.3640 \\
\midrule
\multirow{7}{*}{$\alpha=99\%$}
 & log      & -2.3851 & 3.2098 & 8.3262 & 8.9159 & 9.8836 & 10.9717 \\
 & power03  & -1.8206 & 2.6868 & 6.8816 & 7.3994 & 8.2324 & 9.2742 \\
 & arcsinh  & -2.4234 & 3.2380 & 8.3990 & 8.9908 & 9.9470 & 11.0468 \\
 & arctan   & -1.6265 & 2.9416 & 7.0620 & 7.5615 & 8.3318 & 9.1727 \\
 & rational & -1.3903 & 2.7987 & 6.6296 & 7.1344 & 7.9460 & 8.8084 \\
 & arcsin   & -1.4748 & 2.7397 & 6.5230 & 7.0171 & 7.8315 & 8.7667 \\
 & \emph{static} & 0.2464 & 4.0975 & 6.9995 & 7.6007 & 8.4820 & 9.4159 \\
\bottomrule
\end{tabular}
\caption{Relevant statistics of the terminal hedging error distribution on the $10,\!000$-path test set with initial state perturbation for the set of dynamic risk models trained under the different scoring function characterizations and a static (see last row of each subpanel) risk model across all confidence levels at the one-year maturity. The CVaR statistic is applied to losses (i.e. minus the final P\&L).}
\label{tab:HedgeStats}
\end{table}

%The dynamic risk model achieves nearly optimal hedging performance across all $\alpha$ levels.
From the table and figure, we observe that dynamic risk models yield, in most cases, lower standard deviations, corresponding to narrower P\&L distributions, and lower average P\&L. Indeed, optimizing for a dynamic risk measure fundamentally means minimizing future risk recursively across each conditional state along a path, which entails applying tighter risk control (and higher hedging costs) than what the static risk agent achieves by simply minimizing risk on paths that lead to high risk on final portfolio values. For the $\alpha=92.5\%$, $\alpha=95\%$, and $\alpha=97.5\%$ confidence levels, the static risk models achieve uniformly lower static risk than the dynamic risk models across all scoring functions. This is expected since the static risk agent directly aims to minimize the terminal hedging risk, unlike the dynamic risk agent. 
%\fredcomment{This is a bit worrisome; it means that the static agent did not reach optimality, i.e. poor training?}
%\Shuyicomment{Not necessarily; static policies in my training converged fairly well. I also observed in~\cite{MarzbanDelageLi2023} where the static risk model had slightly worse static expectile risk (2.43) vs. the dynamic risk model (2.38) at the same one-year maturity the models were trained on. Additionally, the superiority of the DRM to SRM in our results is directly correlated with the power03/log/arcsinh functions, which likely pair well with the shifting procedure leading to better DRM policies then expected.}
%\fredcomment{Ok let's discuss I might be missing something.}
Additionally, for the moderate $\alpha=95\%$ and $\alpha=97.5\%$ confidence levels, the dynamic risk models trained under the unbounded sublinear scoring functions achieve static risk results closer to the static risk models than those trained under the saturating scoring functions (i.e. are superior). On the other hand, at $\alpha=92.5\%$ and $\alpha=99\%$, the dynamic risk models trained under saturating scoring function outperforming those trained under unbounded sublinear scores. We may first remark that these results are consistent across scoring functions within each category, suggesting that the observed differences reflect systematic effects of the scoring function classes rather than random variation. The aforementioned behavior can be explained by the increased sensitivity of the unbounded sublinear scoring functions to tail samples. Thus, for moderate $\alpha$ the tail sensitivity of the scoring functions with non-saturating gradients are informative  across the range of cost random variables to the scoring value, thus leading to a better final policy. However, for the slightly lower $\alpha=92.5\%$ confidence level, the tail-sensitivity is less important, favoring a scoring function with bounded and stable updates, explaining the superiority of saturating scoring functions. Interestingly, despite $\alpha=99\%$ placing greater emphasis on the extreme tail, it does not exhibit the same pattern. We posit that at the $\alpha=99\%$ confidence level, sensitivity to tail samples leads to sparse, extreme tail samples which lead to large, disproportionate gradients that destabilize training. In this setting, we can observe that the saturating scoring functions may be preferred even when samples are deep in the tail since their lower sensitivity stabilizes training. This suggests that saturating scoring functions may occasionally be able to handle the variance of sampling for the critic at high confidence levels. The training instability from sparse training samples may also help explain the subpar performance of the $\alpha=99\%$ confidence level static risk model.

% Lastly, one may notice that for both the dynamic and static risk models, the std and CVaR on the terminal P\&L distributions increase when training under greater $\alpha$. We attribute this to the decrease in 
\begin{table}[htbp]
\centering
\small
\begin{tabular}{llrrrrr}
\toprule
 & & \multicolumn{5}{c}{Maturity (months)} \\
Confidence level & Model & 10 & 8 & 6 & 4 & 2 \\
\midrule
\multirow{7}{*}{$\alpha=92.5\%$}
 & \emph{static} & 3.7271 & 3.7837 & 3.9054 & 4.1066 & 4.1438 \\
 & log      & 4.0792 & 3.7239 & 3.4592 & 3.3789 & 3.3546 \\
 & power03  & 4.0894 & 3.7582 & 3.4799 & 3.4031 & 3.3667 \\
 & arcsinh  & 4.0879 & 3.7413 & 3.4783 & 3.3895 & 3.3622 \\
 & arctan   & 3.9358 & 3.5698 & 3.3201 & 3.3035 & 3.3280 \\
 & rational & 3.8212 & 3.4675 & 3.2391 & 3.2631 & 3.3128 \\
 & arcsin   & 3.9114 & 3.5574 & 3.3010 & 3.3005 & 3.3323 \\
\midrule
\multirow{7}{*}{$\alpha=95\%$}
 & \emph{static} & 4.1585 & 4.2206 & 4.3512 & 4.5963 & 4.6317 \\
 & log      & 4.3425 & 3.9402 & 3.6360 & 3.5778 & 3.5833 \\
 & power03  & 4.3431 & 3.9453 & 3.6579 & 3.5873 & 3.5804 \\
 & arcsinh  & 4.3365 & 3.9323 & 3.6429 & 3.5927 & 3.5979 \\
 & arctan   & 4.8387 & 4.3670 & 3.9827 & 3.8107 & 3.6786 \\
 & rational & 4.7777 & 4.3032 & 3.9322 & 3.7696 & 3.6636 \\
 & arcsin   & 4.6820 & 4.2072 & 3.8608 & 3.7232 & 3.6533 \\
\midrule
\multirow{7}{*}{$\alpha=97.5\%$}
 & \emph{static} & 4.8661 & 4.9042 & 5.0659 & 5.3495 & 5.4316 \\
 & log      & 5.8591 & 5.2388 & 4.6610 & 4.3990 & 4.2534 \\
 & power03  & 5.6085 & 4.9661 & 4.4811 & 4.2425 & 4.1644 \\
 & arcsinh  & 5.7868 & 5.1715 & 4.6050 & 4.3461 & 4.1980 \\
 & arctan   & 6.3376 & 5.8142 & 5.2437 & 4.8519 & 4.4383 \\
 & rational & 6.1482 & 5.6704 & 5.1249 & 4.7424 & 4.3695 \\
 & arcsin   & 6.1890 & 5.6750 & 5.1459 & 4.7434 & 4.3813 \\
\midrule
\multirow{7}{*}{$\alpha=99\%$}
 & \emph{static} & 9.1110 & 8.8551 & 8.7340 & 8.1476 & 7.6538 \\
 & log      & 10.0580 & 9.1002 & 7.8508 & 6.8589 & 5.8203 \\
 & power03  & 8.5414 & 7.6511 & 6.6518 & 5.9585 & 5.3270 \\
 & arcsinh  & 10.1246 & 9.1652 & 7.9251 & 6.9072 & 5.8456 \\
 & arctan   & 8.5489 & 8.0814 & 7.1171 & 6.4386 & 5.5849 \\
 & rational & 8.1729 & 7.6108 & 6.7395 & 6.1424 & 5.3965 \\
 & arcsin   & 8.0440 & 7.5185 & 6.6083 & 6.0087 & 5.3163 \\
\bottomrule
\end{tabular}
\caption{Static $\mathrm{CVaR}_{95\%}$ over measurement horizons from 10 to 2 months at 2-month intervals, obtained by evaluating the dynamic risk model under each scoring function and the static risk model baseline at each shorter maturity over a $10,\!000$-path test set with initial state perturbation, across all confidence levels.}
\label{tab:maturityeval}
\end{table}
\begin{figure}[h]
    \centering
    \includegraphics[width=\linewidth]{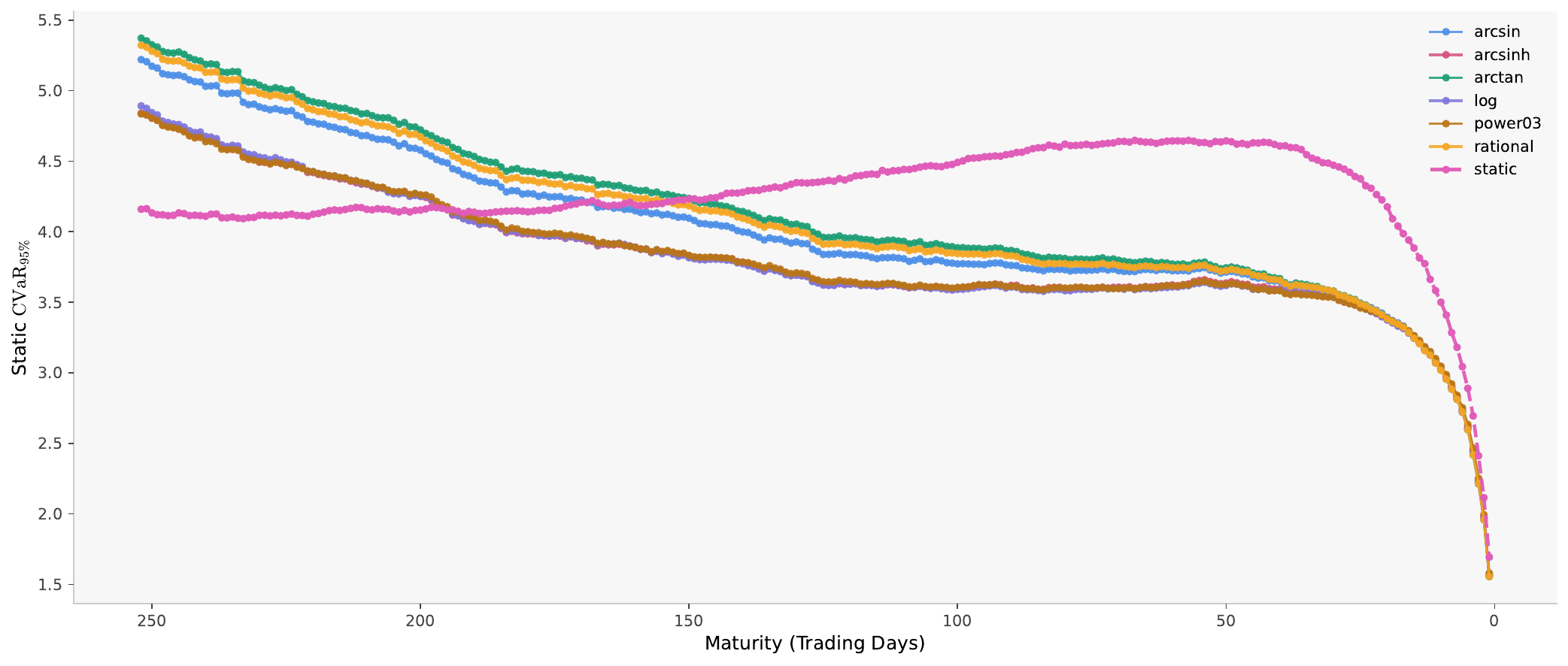}
    \caption{Static $\mathrm{CVaR}_{95\%}$ of the terminal hedging error for the dynamic risk model under each scoring function and for the static risk model baseline, evaluated over a $10,\!000$-path test set with initial state perturbation for each maturity from one year (252 days) to one day.}
    \label{fig:riskovermaturities}
\end{figure}

To further analyze the validity of the dynamic risk models, we adopt the methodology in \cite{MarzbanDelageLi2023} where dynamic risk and static risk models are evaluated according to a static risk metric over horizons that are shorter than the full hedging horizon on which hedging agents are optimized. Similar to their approach, we assume the same initial conditions as the training set but simulate $10,\!000$-path test sets with initial state perturbation to a shorter maturity for performance evaluation of agents trained on longer maturities. 
In~\Cref{fig:riskovermaturities}, we show the hedging performance of the dynamic and static risk models at the $\alpha = 95\%$ confidence level. We report the specific hedging risk values over shorter risk measurement horizons for all confidence levels in~\Cref{tab:maturityeval}. Over shorter horizons from 2 to 6 months, we see that the dynamic risk models exhibit lower static $\mathrm{CVaR}_{95\%}$ than the static risk models over horizons shorter than the full option maturity. This highlights the effectiveness of the time-consistent approach: despite being optimized according to a dynamic risk objective, the resulting models achieve superior static risk performance than the static risk baseline when measuring risk on a horizon shorter than the training horizon. Moreover, we notice that shorter-horizon static risk models display increasing static risk over shorter horizons in contrast to the smooth decrease for dynamic risk models. Indeed, the static risk model becomes increasingly suboptimal (from a local risk perspective) at longer horizons due to time inconsistency, and it incurs greater risk as maturity approaches. %Thus, shorter horizons measured from time $t=0$ should also exhibit greater risk, which can explain this discrepancy. %The time-varying correlations and volatilities of the assets may further contribute to this effect, as the conditional market state can deviate substantially from its unconditional level over the hedging horizon. Overall, the dynamic risk policy does not suffer from this issue because it is optimized at each intermediate time period to remain optimal under the conditional dynamics of the DCC-GARCH model, whereas the static risk policy is optimized once over the full horizon and does not re-optimize the remaining hedge conditional on the realized regime. 

Again, we observe that the relative performance of the dynamic risk models trained under unbounded sublinear scoring functions in comparison to those trained with saturating scoring functions over all confidence levels in~\Cref{tab:HedgeStats} seems to be preserved when evaluated across shorter maturities. That is, in~\Cref{tab:maturityeval}, hedging policies trained under the unbounded sublinear scoring functions continue to be the best performers for $\alpha=95\%$ and $\alpha=97.5\%$ while those trained under the saturating scoring functions continue to be the best performers for $\alpha=92.5\%$ and $\alpha=99\%$. In combination, the consistent differences across scoring-function categories indicate that the relative performance of final hedging policies depend on the characterization of the scoring functions and the tail focus of the risk measure, with different categories (i.e. unbounded sublinear and saturating) performing best at different confidence levels.

Beyond policy performance, we examine the dynamic risk estimates produced by the learned critics. In~\Cref{tab:drm_srm_ratio_t0}, we show the RL-based critic dynamic risk estimate of dynamic risk models across scoring functions and the corresponding static risk of the static risk model baseline over all confidence levels on the $10,\!000$-path test set with the fixed initial state. Additionally, we train separate critics on the final dynamic risk policies trained with the log and rational scoring functions at the $\alpha=95\%$ confidence level to estimate the value function in~\Cref{eq:value function-nested} using the nested approach in~\cite{CoacheJaimungal2024} and gauge the accuracy of the risk estimates from the elicitability-based approach. The neural networks for each time period are trained simultaneously and regress to the dynamic risk from the nested approach using the mean squared error (MSE) Loss.
%\fredcomment{Maybe explain, MSE over what?}
%\Shuyicomment{I think how much the agent should "charge" is the dynamic risk value when $B_0=0$, based on other papers. We could probably use different terminology to avoid possible scrutiny, though, since the equal-risk price would probably be much different}

%First, we may notice that the hedging agent should charge much more under a dynamic $\mathrm{CVaR}_{\alpha}$ objective than the static $\mathrm{CVaR}_{\alpha}$ objective, given that the ratios in the third columns are all significantly greater than $1$. This can be explained by the fact that the dynamic CVaR represents a recursive and nested tail risk-measure, resulting in an accumulation of risk over time. 
We first see that the dynamic risk to static risk ratio increases for greater $\alpha$, indicating that the dynamic CVaR becomes increasingly more conservative relative to its static counterpart as the focus shifts toward more extreme tail losses. %It may be of interest to find the $\alpha$ where the dynamic CVaR at time $t=0$ is equivalent to the static CVaR; this could be an area of future work to better compare time-consistent vs. precommitment frameworks. 
Moreover, we observe a difference in the predicted dynamic risk values between models trained with unbounded sublinear and saturating scoring functions, with the latter consistently yielding lower risk estimates over the final (different) policies over all confidence levels. From a theoretical perspective, this discrepancy can be explained by the fact that saturating scoring functions exhibit weaker signal for extreme pathwise losses at moderate confidence levels, potentially leading to an underestimation of risk and a less accurate baseline for the actor to learn from. Indeed, we find that the RL-based critic trained under the log scoring function yields a dynamic risk estimate of $137.47$ in comparison to the dynamic risk estimate of $150.20$ from the nested approach; in contrast, the RL-based critic trained under the rational scoring function yields a dynamic risk estimate of $108.21$ in comparison to the dynamic risk estimate of $153.82$ from the nested approach. 

Importantly, the previous finding (see \cref{tab:maturityeval}) that hedging policies trained under saturating scoring functions at moderately high $\alpha$ exhibit greater hedging risk over shorter-horizon maturities  than policies trained under the unbounded sublinear scoring functions is consistent with the numerical observation of the greater nested approach critic dynamic risk estimate when the agent is trained under the rational scoring function instead of the log scoring function. 
%, and the RL-based critic estimates lower dynamic risk for the former case than the nested critic. This provides evidence that
Indeed, the lower dynamic-risk estimate produced by the RL-based critic under the rational scoring function reflects an underestimation of the true dynamic risk, while logarithmic-style scoring functions provide a more accurate assessment of dynamic risk and consequently lead to a better policy. Although nested-critic estimates are not obtained for the remaining scoring functions, the consistently lower RL-based dynamic-risk estimates and inferior hedging performance observed under saturating scoring functions, at moderate $\alpha$, together with the more accurate dynamic-risk estimate and stronger hedging performance obtained under the log scoring function, suggest that highly tail-sensitive, non-saturating scoring functions may provide more reliable dynamic-risk estimation in complex risk-aware RL.

%\fredcomment{Make parallel with table 1, higher risk in table 1 = higher risk in table 4.}
%\Shuyicomment{I trained two critics on the log and rational scoring function final policies at $\alpha=95\%$ using the nested approach (using a small batch) just to validate the underprediction claim and show that the policies with saturating scoring functions genuinely have worse estimates of dynamic risk instead of the possibility of lower dynamic risk arising from a better final policy}
% \Shuyicomment{Will need to add some sort of baseline to show underprediction though; since the policy could be lower risk, though this is unlikely since the saturating scores perfromed worse across the maturities.}
\begin{table}[htbp]
\centering
\small
\begin{tabular}{llrrr}
\toprule
$\alpha$ & Scoring & RL-based Dynamic $\CVaR_{\alpha}$ & Nested Dynamic $\CVaR_{\alpha}$ & RL-DR / SR \\
\midrule
\multirow{6}{*}{92.5\%}
& log      & 115.15 & --          & 7.20 \\
& power03  & 102.42 & --          & 6.40 \\
& arcsinh  & 115.20 & --          & 7.20 \\
& arctan   & 79.46  & --          & 4.97 \\
& rational & 77.10  & --          & 4.82 \\
& arcsin   & 77.93  & --          & 4.87 \\
\midrule
\multirow{6}{*}{95\%}
& log      & 137.47 & $\mathbf{150.20}$ & 8.38 \\
& power03  & 125.43 & --          & 7.65 \\
& arcsinh  & 137.83 & --          & 8.41 \\
& arctan   & 110.84 & --          & 6.76 \\
& rational & 108.21 & $\mathbf{153.82}$ & 6.60 \\
& arcsin   & 108.01 & --          & 6.59 \\
\midrule
\multirow{6}{*}{97.5\%}
& log      & 203.22 & --          & 11.90 \\
& power03  & 181.53 & --          & 10.63 \\
& arcsinh  & 202.72 & --          & 11.87 \\
& arctan   & 177.11 & --          & 10.37 \\
& rational & 172.37 & --          & 10.09 \\
& arcsin   & 172.52 & --          & 10.10 \\
\midrule
\multirow{6}{*}{99\%}
& log      & 396.67 & --          & 18.53 \\
& power03  & 347.12 & --          & 16.21 \\
& arcsinh  & 398.03 & --          & 18.59 \\
& arctan   & 223.30 & --          & 10.43 \\
& rational & 216.50 & --          & 10.11 \\
& arcsin   & 216.59 & --          & 10.12 \\
\bottomrule
\end{tabular}
\caption{The dynamic risk estimate at $t=0$ of the dynamic risk model (first column) by passing the fixed initial state to the RL-based critic and its ratio to the corresponding static $\mathrm{CVaR}_{\alpha}$ value of the static risk model evaluated over the $10,\!000$-path test set with a fixed initial state. Since training used an initial account value equal to $\mathcal{V}_0$, both risk objectives were evaluated on the terminal outcome net of this initial value. Since there is also a risk-free interest rate in our hedging problem, we add $\mathcal{V}_0 e^{\bar{r}_f T}$ back to each risk estimate (since CVaR is translation invariant) to obtain the risk under $\mathbb{C}_0 = 0$. We also train a critic using the nested approach from \cite{CoacheJaimungal2024} on the final policies trained under the log and rational scoring functions at the $\alpha=95\%$ confidence level using a smaller test set. 
%by generating an additional $M=2048$ transition states for each state from a batch of a $B=1024$-path test state with a fixed initial state. 
}
\label{tab:drm_srm_ratio_t0}
\end{table}

%\fredcomment{Table (4)   (ii) in the table notes, first sentence, we mention the 10k paths with fixed initial state for the static risk. But for the dynamic risk we also use the fixed states? }
%\Shuyicomment{So the dynamic risk estimate is just passing the fixed initial state to the critic, so there's no test set in this sense. The test set for the static from that same intial state, over $B=1000$ paths, is needed to calculate the static CVaR of the terminal P\&Ls. Yes, I'm okay with removing the description of nested in the description and the things in the other paragraphs.}

% Another observation from the table and figure is that as the dynamic and static  risk models become more risk-averse, the P\&L objectivedistribution becomes counterintuitively wider, differing from the expectation of a narrower distribution as the risk-averse objective means the agent is able to mitigate the risk of extreme outcomes leading to less downside losses and thinner tails.

%As mentioned prior, the precommit strategy is time-consistent w.r.t. a sequence of CVaRs with different confidence levels, and finding these risk measures could be further explored in later work. 

\begin{figure}[t]
    \centering
    \includegraphics[width=\linewidth]{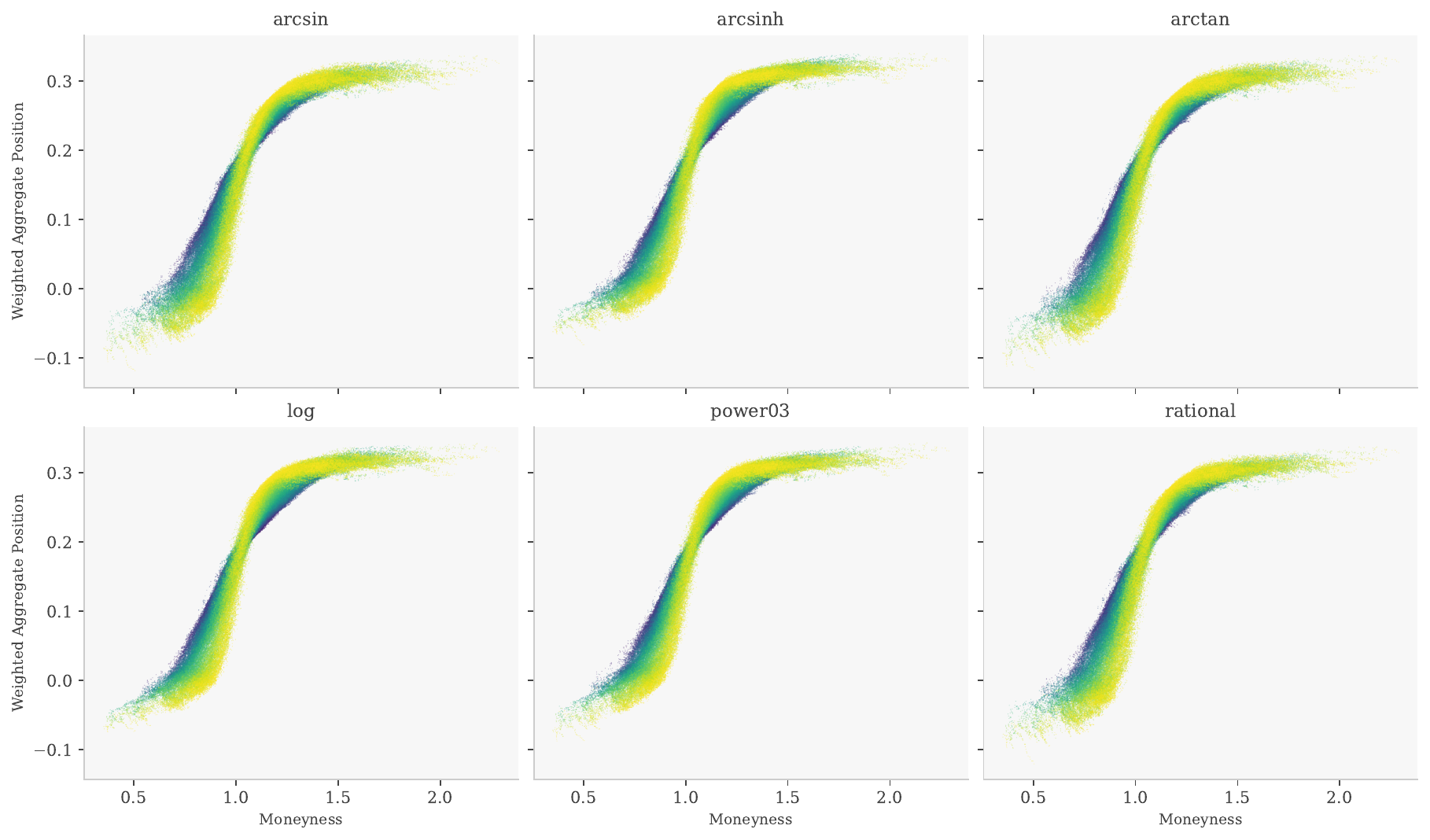}
    \caption{Comparison of the weighted aggregate position of the hedging policy across basket values for the dynamic risk model at the $\alpha = 95\%$ confidence level across the $6$ scoring functions tested across the $10,\!000$-path test set without randomization over all confidence levels. Lighter colors correspond to hedging decisions made closer to the maturity of the basket option.}
    \label{fig:PortDelt}
\end{figure}

Next, we analyze the behavior of the time-consistent policies over time. \Cref{fig:PortDelt} presents a scatterplot of the weighted aggregate position (basket delta), computed as the sum of each asset's position weighted by its corresponding basket weight, against moneyness across all time periods and states in the $10,\!000$-path test set with a fixed initial state, which allows us to visualize the characteristic shape of the delta curve for an initially at-the-money (ATM) basket option. Across all scoring functions, the scatterplot patterns are qualitatively similar and follow the signature delta curve for the writer of an option, validating the reliability of all scoring functions in guiding the actor to learn optimal hedging policies.

\begin{figure}[t]
    \centering
    \makebox[\linewidth][c]{%
        \includegraphics[
            width=1.15\linewidth,
            height=0.82\textheight,
            keepaspectratio
        ]{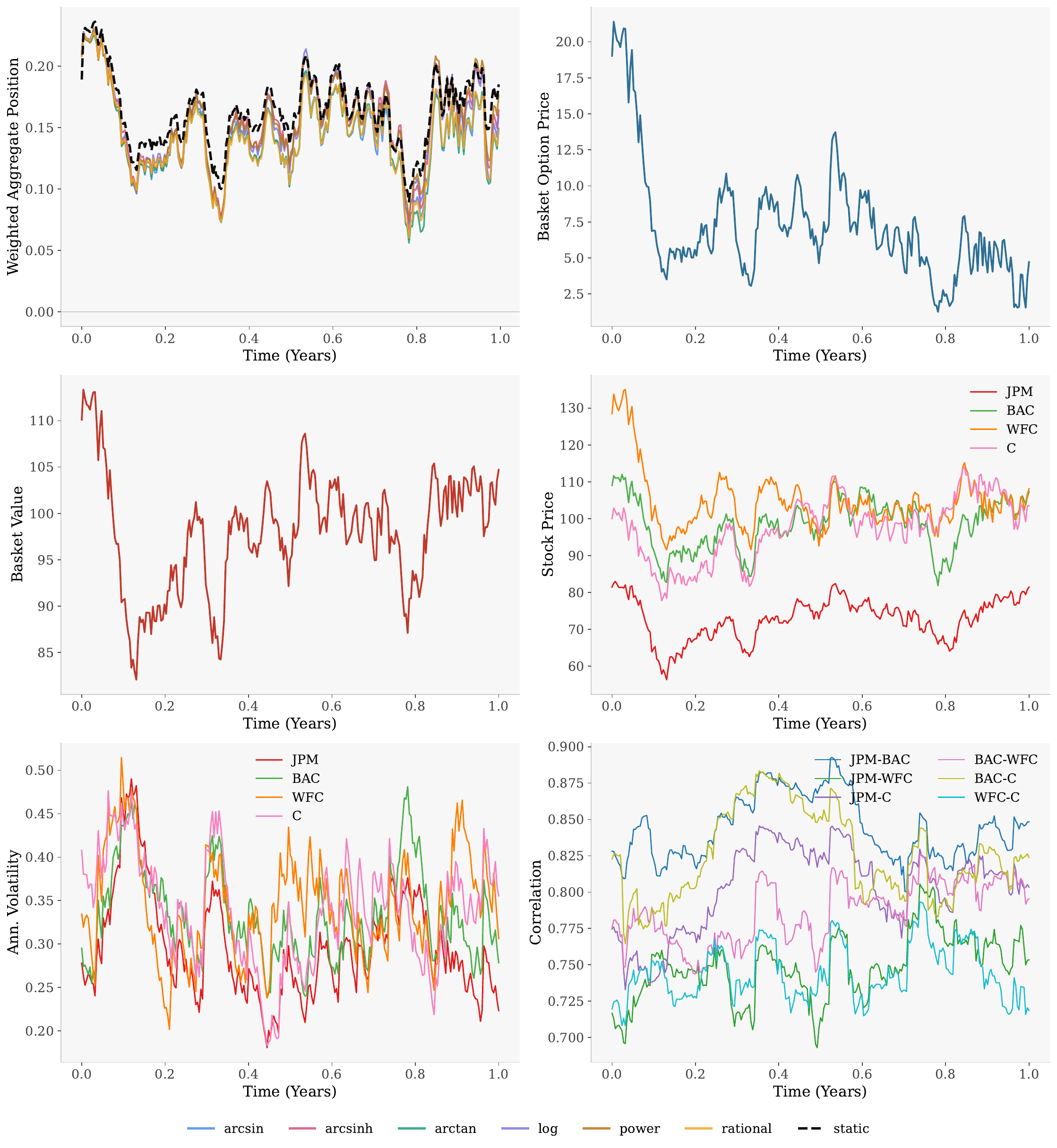}%
    }
    \caption{Comparison of the weighted aggregate position of the dynamic and static risk hedging policies trained at the $\alpha=95\%$ confidence level for JPM, WFC, BAC, and C across the path where the basket option has large daily price changes and exhibits high gamma near expiry from the $10,\!000$-path test set with initial state perturbation.}
    \label{fig:Extremetraj}
\end{figure}

~\Cref{fig:Extremetraj} shows the hedging behavior of the dynamic risk vs. static risk models at the $\alpha=95\%$ confidence level for a trajectory with high-gamma in order to examine how policies hedge in settings with rapid option delta changes and high P\&L variance. The first panel shows the weighted aggregate position (basket delta) across the path for the dynamic and static risk models. The second panel shows the predicted option price using the basket surrogate neural network over the path. The third panel represents the value of the basket and the remaining panels show the price, conditional variance, and pairwise correlations for that specific path. We observe that the time-consistent dynamic risk model generally allocates less than the static model to assets in order to reduce risky exposure to adverse outcomes along the path. This can be explained by the static model attempting to reap the equity risk premium and use it as a buffer against terminal hedging shortfalls. Though desirable from a terminal risk standpoint, this has the effect of generating higher volatility throughout the path as explained above. This explains why dynamic risk optimization refrains from seeking to capture risk premium at the expense of higher local risk.

\begin{figure}[h]
    \centering
    \makebox[\linewidth][c]{%
        \includegraphics[
            width=1.25\linewidth,
            height=0.82\textheight,
            keepaspectratio
        ]{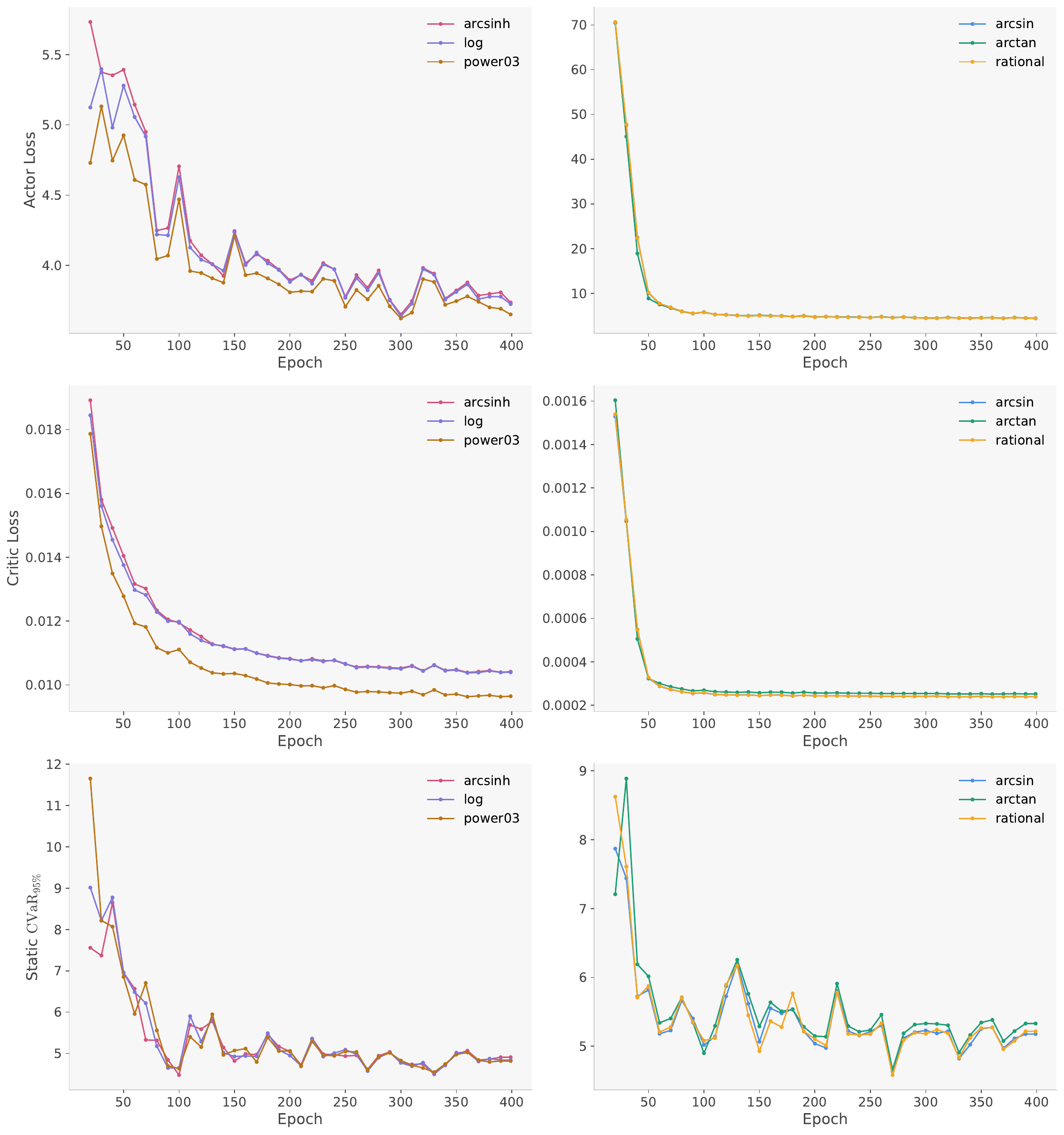}%
    }
    \caption{Plots of the actor loss, the critic loss and measured static test risk over intermediate test sets (see the text for explanations) with plots separated based on type of scoring function for better visualization.}
    \label{fig:losses}
\end{figure}

To assess the convergence and stability of the actor-critic training procedure, we plot the actor loss, critic loss, and test-set (static) risk over the training epochs for the dynamic risk model at the $\alpha=95\%$ confidence level in~\Cref{fig:losses}. These quantities are evaluated every 10 epochs, with the risk computed on separate, intermediate, test sets each comprised of a batch size of $2048$. Such intermediate test sets are reserved for monitoring during training and are distinct from the $10,\!000$-path test sets used for the results in the figures and tables.
The set of left plots represent the unbounded sublinear (log, power03, and arcsinh $G$s) scoring functions while the right plots represent the saturating (arctan, rational, and arcsin $G$s) scoring functions. Visually, we can observe that the actor and critic losses are able to converge to relatively stable values towards the tail-end of training. However, the actor-critic RL algorithm exhibits unstable and slightly worse static risk performance on the intermediate test sets
towards the tail end of training when using saturating scoring functions. In tandem with our prior observations and consistent with the literature, this suggests using unbounded sublinear scoring functions when the risk confidence level considered is moderate.

We remark upon the advantages of the novel shifting procedure introduced in~\Cref{sec:SequentialShifting}, where the value function is estimated sequentially and shifted periodically through the use of an ensemble of neural networks (i.e. one neural network per time period) under the elicitability-based approach. First, in relatively low-dimensional RL settings, using a single neural network while augmenting state vectors to include the time feature instead of using an ensemble to estimate the value function is often the most desirable choice; using a single neural network requires less parameters than an ensemble and often increases the efficiency of training. However, in more complex, high-dimensional settings, the value function may become more complex and differ substantially even across a few time periods, meaning that a single, sufficiently large, neural network with state augmentation may not be able to stably approximate the value function accurately under the elicitability approach. Conversely, an ensemble of smaller neural networks where each neural network attempts to approximate the value function on distinct time periods may be able to more accurately approximate complex value functions in a stable fashion. 
This can materialize since the gradients of the critic loss (see \Cref{eq:criticlossinterm}) are applied separately to each unique parameter set pertaining to each network,
and thus there is no interference in gradient-based updates when refining parameters impacting the critic estimates for multiple time points at once; indeed such updates could pressure moving parameters in opposite directions.
This can lead to a more effective descent of the score landscape for each value function estimation problem at each time period. Second, the use of an ensemble of neural networks means that the value function can learn off of frozen target random variables in stable fashion and be periodically shifted, which ensures that the critic focuses on state-dependent variation rather than the scale and magnitude of the value function over time. 
Indeed, the curvature of the optimization landscape of the elicitability-based approach is significantly more desirable when the random variables have smaller realizations due to the flattening convexity of the scoring function for large tail samples. From our numerical experiments, we find that the actor-critic RL algorithm converges to an optimal policy with accurate estimations of the value function only when using both (i) an ensemble of neural networks, with one network corresponding to each time period, and (ii) the sequential shifting training procedure. In contrast, the actor-critc algorithm tends to diverge (i.e. the loss function blows up) under the following alternatives: (i) when the ensemble of networks is trained simultaneously without sequential shifting, (ii) when the ensemble is trained in sequence in a backwards fashion without shifting, and (iii) when a single neural network is used for all groups while applying shifting (i.e., $\mathbb{G} = 0$). 
%and $\mathbb{G}+1=18$, and when sequential training (i.e. training the groups of neural networks in a backwards fashion) is performed without shifting the value function and vice versa.
%\fredcomment{Here what does vice versa refers to?}
 This implies that our proposed shifting approach is key to successful training in our high-dimensional setting, and that such approach is only fruitful when using separate neural networks for each time step.

\section{Conclusion}
\label{sec:conclusion}
In this paper, we are able to show that an actor-critic RL framework inspired from \cite{CoacheJaimungalCartea2023} can successfully solve high-dimensionality problems; we were able to tackle the hedging of a basket option across much more granular time grids and higher confidence levels than previously proposed in literature. We nevertheless highlight that a value-function translation approach that leverages regions of the scoring function with higher convexity facilitates numerical optimization and is key to obtain stable training results. Moreover, we find that unbounded sublinear (e.g. logarithmic and fractional-power) characterizations of the scoring function for CVaR generally yield final hedging policies with the best performance when sufficient samples are available in risk-aware actor-critic RL. At extreme confidence levels (e.g., $\alpha=99\%$), however, saturating scores may yield better hedging performance when tail samples are scarce, as their diminishing tail sensitivity limits the influence of individual samples and can make the loss more stable. Using the proposed elicitability-based approach, we derive effective dynamic risk models under the CVaR risk measure and evaluate their hedging performance against precommitment policies. While precommitment policies outperform on terminal static risk and achieve greater average terminal P\&L, dynamic risk policies produce less variable terminal P\&L distributions and exhibit superior hedging performance relative to the static risk model over shorter maturities. Overall, these results demonstrate the effectiveness of the elicitability-based approach in RL while highlighting the importance of tuning the scoring function and the optimization landscape for stable learning at high confidence levels.

%Future work should extend time-consistent dynamic risk environments to other applications or actor-critic RL under time-consistent dynamic risk measures, including convergence to optimal policies.
%\fredcomment{These are vague recomm... Do we need them?}
%\Shuyicomment{I will try to think of more specific ones.}

%Our RL agent trained under time-consistent dynamic CVaR maintains comparable but not superior performance to the static risk model, revealing the potential benefit of static risk objectives and precommitment objectives for tail risk mitigation.
%\fredcomment{Maybe revisit that. Not superior in what sense.}
% We also show that all our tested scoring functions for CVaR are appropriate 
%\fredcomment{How do we conclude these are appropriate?}
% given sufficiently convex real-valued functions for our RL experiment, those real-valued functions of $-\log(x)$, $ x^{p} \text{ where } p < 1$, and $-\mathrm{arcsinh}(x)$ exhibit the most accurate characterization of risk and would likely be the best choices in future conditionally elicitable RL frameworks. 
% Suggestions for future research 

%%%%%%%%%%%%%%%%%%%%%%%%%%%%%%%%%%%

\bibliographystyle{apalike}
%\bibliographystyle{abbrvnat}
%\bibliographystyle{unsrtnat}
% plain
\bibliography{references}

%%%%%%%%%%%%%%%%%%

\appendix
\crefalias{section}{appendix}

%%%%%%%%%%%%%%%%%%%%%%%%%%%%%%%%%%%%%%%%%%%%%%%%%%%%%%%%%%%%%%%%%%%%%%%%%%

%%%%%%%%%%%%%%%%%%%%%%%%%%%%%%%%%%%%%%%%%%%%%%%%%%%%%%%%%%%%%%%%%%%%%%%%%%

\section{Hessian Derivation}
\label{app:Hessian Derivation}
%Derivation for the hessian of the score in~\Cref{Cor:ElicitabilityCVaR}. 
We derive \eqref{eq:theHessian}, the Hessian of $\mathfrak{s}(\mathfrak{a}_1, \mathfrak{a}_2) = 
\mathbb{E}_{Y \sim F}[S(\mathfrak{a}_1, \mathfrak{a}_2, Y)]$ with 
respect to $(\mathfrak{a}_1, \mathfrak{a}_2)$, evaluated at the minimizer $\mathfrak{a}_1^* = \mathrm{VaR}_\alpha(Y)$ and 
$\mathfrak{a}_2^* = \mathrm{CVaR}_\alpha(Y)$. $S$ is given 
by \eqref{eq:CVaRScore2}.

\begin{proof}
Define
\begin{equation*}
    M(\mathfrak{a}_1,\mathfrak{a}_2, y) := \mathfrak{a}_2 + \frac{1}{1-\alpha}\left(\mathfrak{a}_1\left(\Ind_{y > \mathfrak{a}_1} - (1-\alpha)\right) - y\Ind_{y > \mathfrak{a}_1}\right).
\end{equation*}
Differentiating $\mathfrak{s}$ w.r.t. $\mathfrak{a}_2$, the terms $-G'(\mathfrak{a}_2)$ and $+G'(\mathfrak{a}_2)$ cancel via the product rule, yielding
\begin{equation*}
    \frac{\partial \mathfrak{s}}{\partial \mathfrak{a}_2} = G''(\mathfrak{a}_2)\,\mathbb{E}\left[M(\mathfrak{a}_1,\mathfrak{a}_2,Y)\right].
\end{equation*}
Furthermore,
\begin{align*}
    \frac{\partial \mathfrak{s}}{\partial \mathfrak{a}_1} &= \frac{G'(\mathfrak{a}_2)}{1-\alpha} \frac{\partial \mathfrak{s}}{\partial \mathfrak{a}_1} \mathbb{E} \left[ \left(\mathfrak{a}_1\left(\Ind_{y > \mathfrak{a}_1} - (1-\alpha)\right) - y\Ind_{y > \mathfrak{a}_1}\right) \right]
\\ &= \frac{G'(\mathfrak{a}_2)}{1-\alpha}  \left[ -\mathfrak{a}_1 f(\mathfrak{a}_1) + \left(1-F(\mathfrak{a}_1)\right) - (1-\alpha) -  \underbrace{\frac{\partial \mathfrak{s}}{\partial \mathfrak{a}_1}\int^\infty_{\mathfrak{a}_1} y f(y) dy}_{=- \mathfrak{a}_1 f(\mathfrak{a}_1)} \right]
    \\ &= -\frac{G'(\mathfrak{a}_2)}{1-\alpha}\left(F(\mathfrak{a}_1) - \alpha\right).
\end{align*}
Both above partial derivatives vanish at $(\mathfrak{a}_1^*, \mathfrak{a}_2^*)$ since $F(\mathfrak{a}_1^*) = \alpha$ and $\mathbb{E}[Y\Ind_{Y > \mathfrak{a}_1^*}] = (1-\alpha)\mathfrak{a}_2^*$; both equalities imply that $\mathbb{E}[M(\mathfrak{a}_1^*,\mathfrak{a}_2^*, Y)] = 0$.

Additionally, second-order partial derivatives are given by
\begin{align*}
    \frac{\partial^2 \mathfrak{s}}{\partial \mathfrak{a}_2^2} &= G'''(\mathfrak{a}_2)\,\mathbb{E}\left[M(\mathfrak{a}_1, \mathfrak{a}_2,Y)\right] + G''(\mathfrak{a}_2),
    \\ \frac{\partial^2 \mathfrak{s}}{\partial \mathfrak{a}_1^2} &= -\frac{G'(\mathfrak{a}_2)}{1-\alpha}\,f(\mathfrak{a}_1),
    \\ \frac{\partial^2 \mathfrak{s}}{\partial \mathfrak{a}_1 \partial \mathfrak{a}_2} &= -\frac{G''(\mathfrak{a}_2)}{1-\alpha}\left(F(\mathfrak{a}_1)-\alpha\right).
\end{align*}

Thus, for arbitrary $(\mathfrak{a}_1, \mathfrak{a}_2)$, the Hessian is
\begin{equation}
    \nabla^2 \mathfrak{s}(\mathfrak{a}_1, \mathfrak{a}_2) = \begin{pmatrix} -\dfrac{G'(\mathfrak{a}_2)\,f(\mathfrak{a}_1)}{1-\alpha} & -\dfrac{G''(\mathfrak{a}_2)(F(\mathfrak{a}_1)-\alpha)}{1-\alpha} \\[10pt] -\dfrac{G''(\mathfrak{a}_2)(F(\mathfrak{a}_1)-\alpha)}{1-\alpha} & G'''(\mathfrak{a}_2)\,\mathbb{E}[M(\mathfrak{a}_1,\mathfrak{a}_2,Y)] + G''(\mathfrak{a}_2) \end{pmatrix}.
\end{equation}
Substituting $F(\mathfrak{a}_1^*) = \alpha$ and $\mathbb{E}[M(\mathfrak{a}_1^*,\mathfrak{a}_2^*, Y)] = 0$ at the optimum, the off-diagonal entries vanish and the Hessian reduces to
\begin{equation}
    \nabla^2 \mathfrak{s}(\mathfrak{a}_1^*, \mathfrak{a}_2^*) = \begin{pmatrix} -\dfrac{f(\mathfrak{a}_1^*)\,G'(\mathfrak{a}_2^*)}{1-\alpha} & 0 \\[10pt] 0 & G''(\mathfrak{a}_2^*) \end{pmatrix}.
\end{equation}
% The $(1,1)$ entry satisfies $-\frac{G'(\mathfrak{a}_2^*)}{1-\alpha}f(\mathfrak{a}_1^*) > 0$ by $f(\mathfrak{a}_1^*) > 0$, $1-\alpha > 0$, and $G'(\mathfrak{a}_2^*) < 0$, and the $(2,2)$ entry satisfies $G''(\mathfrak{a}_2^*) > 0$ by strict convexity of $G$. We can also observe that as the Hessian contains strictly positive diagonal entries it is positive definite at $(\mathfrak{a}_1^*, \mathfrak{a}_2^*)$.
%\fredcomment{How come we can say that $f(\mathfrak{a}_1^*) > 0$? Furthermore, why does $G'(\mathfrak{a}_2^*) < 0$? Where do we need the have the Hessian being positive definite?}
%\Shuyicomment{You are right for $f(\mathfrak{a}_1) > 0$. I removed the positive-definite part.}
\end{proof}

\section{Environment}
\label{app:Env}
\subsection{Parameter calibration}
To simulate the asset price behavior of the four bank assets mentioned, we follow the DCC-GARCH(1, 1) model as specified by \citep{Engle2002}, which models individual asset volatilies using a univariate GARCH process and their dependence through time-varying correlations. To capture the tail-dependence across banking assets, we use Gaussian marginals coupled through a Student-$t$ copula with $\nu$ degrees of freedom. For $N$ assets and $t \in {\mathcal{T}_{\text{hist}} = \{1,...,T_{\text{hist}}\}}$, the DCC-GARCH model can be described by first specifying the GARCH dynamics for each asset, followed by the dynamic conditional correlation structure. First, we have
\begin{align}
    \mathbf{r}_t = \boldsymbol{\mu}_t + \boldsymbol{\epsilon}_t, \qquad
    \mathbf{z}_t = D^{-1}_t\boldsymbol{\epsilon}_t, \quad \text{and} \\
    \mathbf{z}_t = \begin{bmatrix}z_{1, t}, \dots,z_{N, t}\end{bmatrix}^{\top},
\end{align}
where $\mathbf{r}_t$ is the vector of asset returns, $\boldsymbol{\mu}_t$ is the vector of conditional means, and $\boldsymbol{\epsilon}_t$ is the vector of return innovations.  $z_{i,t}$ represents the standardized innovation for the $i$th asset, with a standard normal marginal distribution. $D_t$ is the diagonal matrix of conditional volatilities.
For our hedging simulation, each asset follows a Heston--Nandi 
GARCH(1,1) process \citep{HestonNandi2000} under the physical measure, where 
\begin{equation}
    r_{i,t} = r_{f, t} + \lambda_i h_{i,t} + 
    \sqrt{h_{i,t}}\,z_{i, t}, 
\end{equation}
\begin{equation}
    h_{i,t} = \omega_i + \beta_i h_{i,t-1} + 
    \tilde{\alpha}_i\!\left(z_{i,t-1} - \gamma_i\sqrt{h_{i,t-1}}\right)^2.
\end{equation}
The dependence structure of the innovation vector is driven by a Student-$t$ copula with correlation matrix $R_t$, which
%$D_t R_t D_t$ is the conditional covariance matrix with $R_t$ and $D_t$ being the dynamic conditional correlation matrix and 
%The time-varying correlation matrix $R_t$ 
is updated according to the proxy matrix
\begin{equation}
    Q_t = (1-\alpha_{\text{DCC}}-\beta_{\text{DCC}})\bar{Q}
    + \alpha_{\text{DCC}}\,\boldsymbol{\xi}_{t-1}\boldsymbol{\xi}_{t-1}^{\top}
    + \beta_{\text{DCC}}Q_{t-1},
    \label{eq:DCCupdate}
\end{equation}
where the $\boldsymbol{\xi}_{t-1}$ are defined as follows. First, $\mathbf{u}_t = T_{\nu}^{-1}\left(\Phi\left(\mathbf{z}_t\right)\right)$ is the Student-$t$-transformed standardized residuals, with $\Phi$ denoting the standard normal cumulative distribution function and $T_{\nu}^{-1}$ denoting the quantile function of a standard Student-$t$ distribution with $\nu$ degrees of freedom, and $\boldsymbol{\xi}_t = \sqrt{\frac{\nu-2}{\nu}}\,\mathbf{u}_t$ is their unit variance rescaling (assuming $\nu > 2$). $\bar{Q} = \frac{1}{T_{\text{hist}}}\sum_{t=1}^{T_{\text{hist}}}\boldsymbol{\xi}_t\boldsymbol{\xi}_t^\top$ represents the unconditional proxy matrix where $Q_1 = \bar{Q}$ during calibration. Based on this recursive update, the time-$t$ correlation matrix is then given by
\begin{equation}
    R_t = \operatorname{diag}(Q_t)^{-\frac{1}{2}}\,Q_t\,
          \operatorname{diag}(Q_t)^{-\frac{1}{2}}.
    \label{eq:CCM}
\end{equation}

%\fredcomment{Are the epsilons assumed to be Gaussian? (you would need that to compute the probability-integral-transformed residuals mentioned below)}
%\Shuyicomment{Changed to gaussian}

To estimate the parameters of the DCC-GARCH model, we utilize a two-stage Maximum Likelihood Estimation (MLE) procedure. The data consists of a 10-year historical period from December 23rd, 2015 to December 20th, 2025, containing $T_{\text{hist}} = 2513$ daily closing prices. All market data, including adjusted closing prices for equities and the 3-month Treasury bill yield, are obtained from Yahoo! Finance. The Heston--Nandi GARCH residuals $\epsilon_{i,t}$ are computed recursively for each parameter vector during the MLE procedure from excess log returns by subtracting the continuously compounded risk-free component $r_{f,t}$ and the conditional risk premium $\lambda_{i}h_{i, t}$, which sum to ${\mu}_{i, t}$. $r_{f,t}$ is derived from the annualized 3-month Treasury bill yields and the aforementioned constant daily risk-free rate $\bar{r}_f = \frac{1}{T_{\text{hist}}}\sum_{t=1}^{T_{\text{hist}}} r_{f,t}$. During calibration, the initial variance $h_{i, 1}$ is the sample variance of log-returns. The GARCH parameters $\Theta_i = \{\omega_i,\tilde{\alpha}_i,\beta_i,\gamma_i,\lambda_i\}$ for each asset are estimated by minimizing the following conditional Gaussian negative log-likelihood using the Differential Evolution algorithm:
\begin{equation}
    \min \limits_{\zeta_i} -\sum_t \ell^{\text{Marginal}}_t = \min_{\zeta_i} \frac{1}{2}\sum_t (\ln(2\pi) + \ln(h_{i, t}) + z^2_{i, t}).
\end{equation}

GARCH parameter estimates for the four assets are provided in~\Cref{tab:calibrated_params}.
\begin{table}[h!]
\centering
\caption{Calibrated GARCH parameters}
\label{tab:calibrated_params}
\begin{tabular}{lcccccc}
\hline
Bank & $\omega$ & $\tilde{\alpha}$ & $\beta$ & $\gamma$ & $\lambda$ & $\sqrt{252 \bar{h}}$ \\
\hline
JPM & $4.26\times10^{-8}$ & $1.17\times10^{-5}$ & 0.8668 & 87.19 & 2.4522 & 25.73\% \\
BAC & $1.58\times10^{-7}$ & $1.41\times10^{-5}$ & 0.8563 & 84.94 & 1.3468 & 29.40\% \\
WFC & $1.09\times10^{-9}$ & $2.85\times10^{-5}$ & 0.8590 & 47.39 & 0.6377 & 30.57\% \\
C   & $1.00\times10^{-9}$ & $2.01\times10^{-5}$ & 0.8586 & 65.86 & 0.9214 & 30.54\% \\
\hline
\end{tabular}
\end{table}
Following \cite{HestonNandi2000}, for each asset, the unconditional variance is\footnote{See p.598 of \cite{HestonNandi2000}.} \mbox{$\bar{h}_{i} = (\omega_{i} + \tilde{\alpha}_{i})/(1 - \beta_i - \tilde{\alpha_{i}}\gamma_{i}^2)$} and the annualized unconditional volatility is $ \sqrt{252 \bar{h}_{i}}$. 

Following the multivariate Student-$t$ copula described by \citep{DemartaMcNeil2005} within the conditional copula framework of \citep{Patton2006}, the DCC parameters and degrees of freedom are estimated by minimizing the following negative log-likelihood using the L-BFGS-B algorithm:
\begin{equation}
    \min_{\alpha_{\text{DCC}}, \beta_{\text{DCC}}, \nu} -\sum_t \ell_t^{\text{cop}}
\end{equation}
where
\begin{align*}
    \ell_t^{\text{cop}} &= -\tfrac{1}{2}\log|R_t| 
    - \tfrac{\nu+N}{2}\log\!\left(1 + \frac{\mathbf{u}_t^\top 
    R_t^{-1}\mathbf{u}_t}{\nu}\right)
    + \tfrac{\nu + 1}{2}\sum_{i=1}^{N} \log\!\left(1 + \frac{u_{i,t}^2}{\nu}\right) + \tilde{C},
    \\ \tilde{C} &= \log \Gamma \left(\frac{\nu+N}{2}\right) + (N-1)\log \Gamma \left(\frac{\nu}{2}\right) - N\log \Gamma\left( \frac{\nu+1}{2}\right).
\end{align*}

\begin{table}[h!]
\centering
\caption{Calibrated DCC parameters}
\label{tab:dcc_params_P}
\begin{tabular}{llll}
\hline
Parameter & Value & Constraint & Interpretation \\
\hline
$\alpha_{\text{DCC}}$ 
& 0.0165 & $\geq 0$ 
& News (shock) effect \\
$\beta_{\text{DCC}}$  
& 0.9652 & $\geq 0$ 
& Persistence effect \\
$\alpha_{\text{DCC}}+\beta_{\text{DCC}}$ 
& 0.9816 & $<1$ 
& Stationarity \\
$\nu$ (DoF) 
& 15.9241 & $>2$ 
& Tail thickness \\
\hline
\end{tabular}
\end{table}
The resulting DCC parameters and degrees of freedom are reported in \cref{tab:dcc_params_P}, while the associated unconditional correlation between assets implied by the unconditional proxy $\bar{Q}$ is summarized in \cref{tab:qbar_corr}. Exact calibrated parameter values in our experiments can be found in the configurations folder in the \href{https://github.com/shuyizhang01/Time-consistent_Deep-hedging}{GitHub repository}. 

\begin{table}[h!]
\centering
\caption{Unconditional pairwise correlations from $\bar{Q}$}
\label{tab:qbar_corr}
\begin{tabular}{lc}
\hline
Asset Pair & Correlation \\
\hline
JPM -- BAC & 0.8282 \\
JPM -- WFC & 0.7167 \\
JPM -- C   & 0.7746 \\
BAC -- WFC & 0.7753 \\
BAC -- C   & 0.8227 \\
WFC -- C   & 0.7197 \\
\hline
\end{tabular}
\end{table}
To simulate over $\mathcal{T}$, we generate correlated innovations from the calibrated Student-$t$ copula using the Gaussian scale-mixture representation of the multivariate Student-$t$ distribution. Specifically, at each time period, we draw
\begin{equation}
\label{eqgenuhat}
    \hat{\mathbf{u}}_t
    = \sqrt{\frac{\nu}{\chi_t}}\,\hat{\pmb{\zeta}}_t,
    \qquad
    \hat{\pmb{\zeta}}_t \sim \mathcal{N}(\mathbf{0},R_t),
    \qquad
    \chi_t \sim \chi^2_\nu, 
\end{equation}
where $\hat{\pmb{\zeta}}_t$ and $\chi_t$ are independent. Note that the same $\chi_t$ applies to the denominator for all components of $\hat{\pmb{\zeta}}_t$ in \eqref{eqgenuhat}. Then, $\hat{\boldsymbol{\xi}}_t = \sqrt{\frac{\nu-2}{\nu}}\,\hat{\mathbf{u}}_t$ is used to update $Q_t$ and $\hat{\mathbf{z}}_t = \Phi^{-1}\left( T_{\nu} (\hat{\mathbf{u}}_t)\right)$ is used to update $r_{i, t}$ and $h_{i, t}$. The initial variances and proxy matrix are set to their unconditional values.
%\Shuyicomment{Changed transform to operate on $\mathbf{u}_t$ since  $T^{-1}_{\nu}$ before was assumed to have regular not unit variance. Also, changed $\mathbf{w}_t$ to $\mathbf{u}_t$ because w conflicts with the basket weights and $\mathbf{x}_t$ to $\boldsymbol{\xi}_t$ because $x$ was input to $G$ and some other functions before. Also changed any other conflicting variabels; please feel free to check the changes}
%\fredcomment{I think there might be an issue here; since the $\chi_t$ are generated independently from the $\hat{\pmb{\zeta}}_t$, I don't think the $\hat{\boldsymbol{\xi}}_t$ have a correlation matrix $R_t$... I think the proper approach would be to generate observations jointly from a student-t copula first, and the rescale them to normality (as right now the generated covariates are marginally student, but don't have a student copula dependence).}
%\fredcomment{Check if z are obtained $\mathbf{x}_t = T_{\nu}^{-1}\left(\Phi(\left(\mathbf{z}_t\right)\right)$ I think they should embed dependence as they have not been decorrelated...}
%\Shuyicomment{Sorry; To clarify, I meant to say that we don't know the dependence yet $R_t/Q_t$ when deriving $x_t$. $x_t$ does embed depedence, but they represent the historical residuals mapped to Student-$t$ space. I changed $u_t$ to be $\hat{x}_t$ to represent innovations instead of residuals.}
The Monte Carlo paths used for basket option pricing described in~\Cref{sec:NNP} are generated under the risk-neutral measure. Specifically, following risk-neutral parameterization described in~\cite{HestonNandi2000} and its componentwise extension to the multivariate DCC-GARCH setting, the market price of risk is set to $\lambda_i^{\mathbb{Q}}=-\frac{1}{2}$, and the leverage parameter is adjusted according to
\begin{equation}
    \gamma_i^{\mathbb{Q}} = \gamma_i^{\mathbb{P}} + \lambda_i^{\mathbb{P}} + \frac{1}{2}.
\end{equation}
We assume that the correlation structure of the DCC and the student copula dependence module are left unchanged under the risk-neutral measure.
%\Shuyicomment{Should we say something about we assume DCC copula and correlation dynamics is unchanged, since I don't use option prices to find the corr. or vol. risk premia to make the setup exactly risk-neutral}
% The risk-neutral ($\mathbb{Q}$) DCC parameters in~\Cref{tab:dcc_params_PQ} are therefore model-implied and obtained by \rc{applying adjustments to physical measure parameters}. % using the adjusted GARCH parameters in our MLE.

\subsection{Basket option pricing}
\label{sec:NNP}
To price the European basket call option with a one-year maturity for any $t \in \mathcal{T}$, we use a feedforward neural network described in~\Cref{app:hypBasket} to map the state vector $s_t$ in~\Cref{sec:ExpMeth} excluding $a_{t-1}$ to the time-value of the hedged basket option in a $\logonep(x)$
space to stabilize the time-value's heavy right skew. The training set comprises $n_{\text{samples}} = 100,\!000$ randomized configurations of the state vectors that span the one-year hedging horizon and are stratified across intrinsic-value cases, boundary regimes such as deep ITM/OTM and near-expiry and long-expiry ATM, and a general population from a mixture distribution that oversamples near-expiry maturities, as pricing noise is quite large in the latter regime. 

In the dataset, the response variable for each observation consists of the time value (price minus intrinsic value) of an option price associated with the state $s_t$, which is approximated through Monte Carlo simulation. All Monte Carlo paths in this context are generated according to DCC-GARCH under the risk-neutral measure, implied by the Heston-Nandi transformation described prior.
Each sample (option price) involves applying Monte Carlo simulation with $n_{\text{paths}} = 100,\!000$ paths under the model-implied risk-neutral DCC-GARCH following the same Gaussian marginal and Student-$t$ copula assumption, beginning from a single randomly generated state vector. For each sample, the price of the basket option is calculated as the sample average discounted intrinsic payoff $e^{-\bar{r}_f (T-t)}\,\frac{1}{n_{\text{paths}}}
\sum_{j=1}^{n_{\text{paths}}}
\left[\sum_{i=1}^{N}w_i S_T^{(i,j)}-K\right]_+$ of the $100,\!000$ terminal prices resulting by the DCC-GARCH simulation. The surrogate also prices the zero-time-to-expiry case, for which the option value is given directly by its intrinsic payoff. The network regresses to the time value in the $\logonep(x)$ space using the MSE loss on $85,\!000$ samples until convergence. On a test set of $15,\!000$ samples, the surrogate neural network accurately predicts the time value of the Monte Carlo-priced basket option with a best mean absolute error (MAE) of $\$0.0288$ and median absolute error and $\$0.0150$. Relative to the average 
test time-value of $\$2.1418$, this corresponds to a best relative MAE of 
$1.34\%$. Furthermore, predictions on $95\%$ of test samples exhibit best MAEs below $\$0.0927$.

\section{Neural network architecture and hyperparameters}
\label{app:hyp}
\subsection{Dynamic and static risk models}
\label{app:hypACAO}
As mentioned, there are two neural networks that make up the respective critics for each time period, one estimating VaR and the other estimating the excess between the VaR and CVaR. Each neural network consists of two linear layers of dimension $128$ with the SiLU activation in between, and the excess network applies the LeakyReLU activation to the output with the negative slope parameter set to $.05$. The actor for both the dynamic and static risk models share an identical trunk of two linear layers of dimension 256 and LeakyReLU activations with the negative slope parameter set to $.05$ which feeds into a final linear layer that outputs the mean of the action distribution. The standard deviation of the action distribution for the dynamic risk model is fixed throughout training to $1 \times 10^{-3}$. The actor and critic learning rates for the dynamic risk model are both set to $2.5\times10^{-3}$, with geometric decay rates of $.997$ and $.99999$ per gradient step, respectively. The static risk model's actor and smooth auxiliary variable $q$ have learning rates of $2.5\times10^{-5}$ and $2.5 \times 10^{-2}$, both having a geometric decay rate of $.998$ per gradient step.\footnote{At the $\alpha=99\%$ confidence level, all learning rates are decreased by three orders of magnitude to mitigate instability from sparse, tail-driven updates.} 
For each epoch of the actor-critic algorithm, which involves passing all groups in $\mathcal{G}$ for the critic and updating the actor, the VaR and excess critics for all time periods in each group are updated simultaneously using $E_1 = 2000$ gradient steps for the first epoch and $E_1=1000$ for the rest, and the actor is updated using $E_2=3$ gradient steps. The target neural networks for the critics are updated every $M = 300$ gradient steps.
We set $\mathbb{G}=17$ for 18 total groups of critics over the time horizon for the actor-critic algorithm; we did not test other values for $\mathbb{G}>0$ for the sequential training procedure. The actor-only (static) algorithm takes $1$ gradient step per epoch (i.e. new paths are generated every time the actor is updated). We use $B=16,\!384$, which represents the batch size of generated paths per epoch for critic training, $B_1=2048$ as the minibatch size for critic training, and $B_2=2048$ as the baseline, unadjusted batch size of generated paths for each actor gradient step in both the dynamic and static risk models.~\Cref{tab:risk-level-paths} shows the total generated paths for increasingly risk-averse dynamic risk models for a singular epoch that includes a $B_1$ trajectories for the intermediate test set, $B$ trajectories for the critic, and $B_2/(1-\alpha)$ trajectories for each gradient step of the actor.
%\fredcomment{Might recall here what we mean by an epoch; recall the sequence in word, an epoch = passing across all groups G for the critic + doing actor updates? Difference between outer vs inner epoch.}
%\Shuyicomment{Yes}
%\Shuyicomment{I changed to refer to the outer iteration as an epoch (to avoid confusion with the episode term in RL) and inner iterations as gradient steps.}
%\fredcomment{Explain in Table 8 where does the 3B2/(1-alpha) comes from.}
%\Shuyicomment{Addressed before the pseudocode algo.}
\begin{table}[h]
\centering
\caption{Number of simulated paths per epoch by risk level for the actor-critic RL algorithm}
\label{tab:risk-level-paths}
\begin{tabular}{|c|c|}
\hline
$\alpha$ & \footnotesize{Paths per epoch ($B + B_1 + 3B_2/(1-\alpha)$)} \\
\hline
$92.5\%$ & $100{,}352$ \\
$95\%$  & $141{,}312$ \\
$97.5\%$ & $264{,}192$ \\
$99\%$  & $632{,}832$ \\
\hline
\end{tabular}
\end{table}
The optimizers for the critic and actor for the actor-critic algorithm are Stochastic Gradient Descent with Nesterov Momentum of $.9$ and Adam, respectively. The optimizers for the auxiliary variable and the actor in the actor-only algorithm are both Adam. All dynamic risk models train for $E=400$ outer epochs (equivalent to $1200$ actor updates), while all static risk models train for $E=2000$ epochs (equivalent to $2000$ actor updates). All scoring functions use $C=40$.

For each trial, corresponding to a given scoring function and confidence level, training is conducted on a single NVIDIA T4 GPU with 15 GiB of GPU memory through Kaggle's servers.
\subsection{Basket surrogate}
\label{app:hypBasket}
The Basket Surrogate feedforward neural network contains fully connected layers with widths of $256$, $128$, and $64$, followed by a fully connected output head with dimensions $64$, $32$, and $1$. The network applies layer norm after the first linear transformation, SiLU activations between layers, and dropout regularizations at a rate of $5\%$. A residual connection adds the output of the $128$-unit layer to a learned linear projection of its $256$-dimensional input, with the resulting $128$-dimensional representation passed through dropout and then to the $64$-unit fully connected layer. This is done to achieve effective gradient propagation. To train the surrogate neural network, we use the AdamW optimizer with a learning rate of $1 \times 10^{-3}$ and weight decay of $1 \times 10^{-4}$. A batch size of $512$ is used to train the surrogate model each epoch. If the MSE Loss shows no improvement for a patience period of 20 consecutive epochs, the learning rate is halved until reaching a minimum floor of $1 \times 10^{-6}$. The model trains for $1000$ epochs and stops training early if the latest MAE on the test set has not exceeded its best MAE for $250$ epochs.

\section*{Acknowledgements}
The authors thank Janine Sharbaugh for her invaluable support throughout this work. Frédéric Godin acknowledges funding from NSERC (RGPIN-2024-04593).

\section*{Statement of interest}
The authors have no conflict of interest to declare. 

\end{document}